\documentclass[lettersize,journal]{IEEEtran}
\usepackage{amsmath,amsfonts}
\usepackage{algorithmic}
\usepackage{algorithm}
\usepackage{array}
\usepackage{textcomp}
\usepackage{stfloats}
\usepackage{url}
\usepackage{verbatim}
\usepackage{graphicx}
\usepackage{cite}

\usepackage{amsthm}
\usepackage{braket}
\usepackage{subfigure}
\usepackage{xcolor}
\usepackage{colortbl}

\newtheorem{theorem}{Theorem}
\newtheorem{assumption}{Assumption}
\newtheorem{lemma}{Lemma}

\begin{document}

\title{Quantum Imitation Learning}


\author{Zhihao~Cheng, Kaining~Zhang, Li~Shen,  Dacheng~Tao,~\IEEEmembership{Fellow,~IEEE,}%

\thanks{Manuscript submitted to a journal for review on January 5, 2022.}

\thanks{Zhihao Cheng and Kaining Zhang are with the School of Computer Science, in the Faculty of Engineering, at the University of Sydney, 6 Cleveland St, Darlington, NSW 2008, Australia. E-mails: zche3121, kzha3670@uni.sydney.edu.au.}

\thanks{Li Shen is with the JD Explore Academy, China. E-mail: mathshenli@gmail.com.}

\thanks{Dacheng Tao is with the University of Sydney, NSW 2006, Australia, and also with the JD Explore Academy, China. E-mail: dacheng.tao@gmail.com.}

}

\markboth{Journal of \LaTeX\ Class Files,~Vol.~14, No.~8, August~2021}%
{Cheng \MakeLowercase{\textit{et al.}}: Quantum Imitation Learning}


\maketitle

\begin{abstract}
Despite remarkable successes in solving various complex decision-making tasks, training an imitation learning (IL) algorithm with deep neural networks (DNNs) suffers from the high computation burden. In this work, we propose quantum imitation learning (QIL) with a hope to utilize quantum advantage to speed up IL. Concretely, we develop two QIL algorithms, quantum behavioural cloning (Q-BC) and quantum generative adversarial imitation learning (Q-GAIL). Q-BC is trained with a negative log-likelihood loss in an off-line manner that suits extensive expert data cases, whereas Q-GAIL works in an inverse reinforcement learning scheme, which is on-line and on-policy that is suitable for limited expert data cases. For both QIL algorithms, we adopt variational quantum circuits (VQCs) in place of DNNs for representing policies, which are modified with data re-uploading and scaling parameters to enhance the expressivity. We first encode classical data into quantum states as inputs, then perform VQCs, and finally measure quantum outputs to obtain control signals of agents. Experiment results demonstrate that both Q-BC and Q-GAIL can achieve comparable performance compared to classical counterparts, with the potential of quantum speed-up. To our knowledge, we are the first to propose the concept of QIL and conduct pilot studies, which paves the way for the quantum era. 
\end{abstract}

\section{Introduction}\label{sec:introduction}
\IEEEPARstart{I}{mitation} learning (IL) is originated from the fact that intelligent creatures usually master skills by imitating their parents or peer fellows \cite{meltzoff2013imitation}. In particular, agents are provided with some expert data and required to replicate the expert behaviors recorded by expert data \cite{Bain95}. Empowered by classical deep neural networks (DNNs), IL has been successfully employed in various applications \cite{peng2018deepmimic,codevilla2018end,fang2019survey,li2021automated,pan2020imitation} in the past decades. In general, IL contains two major spectra \cite{hussein2017imitation}, behavioural cloning (BC) and inverse reinforcement learning (IRL). However, current DNNs-based BC and IRL algorithms face a long-standing challenge, i.e., the large sample complexity of training an agent~\cite{sasaki2018sample,blonde2019sample}.

Simultaneously, enormous advances have emerged in quantum computing \cite{gyongyosi2019survey, zhong2020quantum, arute2019quantum}, making it possible to be employed in real-world applications. For example, Google achieved the quantum supremacy in 2019 using a 53-qubit superconducting quantum processor \cite{arute2019quantum}, which performs random circuits sampling exponentially faster than classical computers. Compared to classical counterparts, the proven exponential or quadratic computing speed-up of quantum algorithms for certain complex problems is known as quantum advantage (or quantum supremacy) \cite{biamonte2017quantum, havlivcek2019supervised}. A possible cause of quantum advantage is that quantum systems can process an exponential number of states simultaneously via quantum superposition and entanglement, whereas classical computing could only process one state \cite{biamonte2017quantum} at the same time. Machine learning (ML) is a promising field to apply quantum computing, since it demands huge computing resources \cite{dean2018new}. Some researchers proposed quantum machine learning (QML) with a hope to speed up ML algorithms by quantum advantage \cite{biamonte2017quantum,nguyen2020benchmarking}. Extending the quantum advantage to more general problems and realizing it for practical applications deserve further investigations \cite{cerezo2021variational}.

Although quantum algorithms have been developed for many ML tasks, little attention has been poured to leveraging quantum computing in IL, a research domain that attracts extensive interest and is computation-consuming due to high sample complexity. Compared to reinforcement learning (RL), which has been enhanced with quantum technologies, IL possesses two incomparable merits. First, IL does not require designing a subtle reward function. In some cases, constructing a proper reward function is non-trivial \cite{dulac2020empirical} because it involves background knowledge about the task. Second, IL can utilize existing expert data that are sampled from human beings~\cite{peng2018deepmimic}. For example, there are various online videos that record how experts accomplish tasks \cite{cheng2021guaranteed}, which can be used as expert data for IL. We can obtain efficiency boost with direct guidance from these expert data compared to training from scratch in RL. As a result, we believe that it is significant to conduct pilot studies on fusing potential advantages of quantum technologies and IL.

In this paper, we aim to empower IL with the promising quantum computing and thus propose quantum imitation learning (QIL). Instead of building policies with DNNs, we construct quantum neural networks (QNNs) with variational quantum circuits (VQCs) to represent agents' policies, which are suitable for the current noisy intermediate-scale quantum (NISQ) era. The expressivity of VQCs is enhanced with data re-uploading and scaling parameters in VQCs' inputs and outputs. Classical data are first encoded into quantum states via qubit encoding and then uploaded to VQCs. After processing, we measure quantum observables to read out classical control signals, whose outputs are equipped with a softmax layer to improve exploration in IL. Corresponding to classical IL categories where IL is divided into BC and IRL, we present two QIL algorithms, quantum behavioural cloning (Q-BC) and quantum generative adversarial imitation learning (Q-GAIL). Q-BC learns in an SL manner with a negative log-likelihood (NLL) loss to clone expert behaviors from expert data, which works off-line. By contrast, Q-GAIL maintains a discriminator to generate rewards and then learns to achieve high rewards that are computed by the discriminator, which is on-line and on-policy. To deal with the instability of the adversarial training scheme of Q-GAIL, we adopt spectral normalization to regularize the discriminator and a baseline to reduce variance in optimization of the generator. Our contributions are summarized as follows:
\begin{itemize}
    \item We are the first to improve IL with quantum technologies and propose quantum imitation learning (QIL), paving the way for the coming quantum era.
    \item We construct VQCs for QIL and devise two proof-of-principle QIL algorithms, Q-BC and Q-GAIL, evolving from two spectra of traditional IL algorithms. Q-BC is suitable for the IL problem where there is a large amount of expert data or it is costly to interact with environments to sample new data. Q-GAIL works better under the situation where a limited amount of expert data is provided and it is feasible to sample more data.
    \item Experiment results demonstrate that the proposed QIL algorithms Q-BC and Q-GAIL can achieve expert-level performance in both discrete-action and continuous-action environments for the first time. 
\end{itemize}

The paper is organized as follows. Section \ref{sec:relatedwork} introduces previous works related to IL and QML, while background knowledge such as basic notations and preliminaries is presented in Section \ref{sec:background}. In Section \ref{sec:method}, we construct VQCs for IL and present two QIL algorithms with details, Q-BC and Q-GAIL. Experiments are conducted to demonstrate the performance of QIL algorithms in Section \ref{sec:experiments}. Section \ref{sec:conclusion} concludes the paper and provides the future vision of QIL.

\section{Related Work}\label{sec:relatedwork}

\subsection{Imitation Learning (IL)}
One of the most vital way for intelligent creatures to master skills is by imitating their parents, the elders, or even peer fellows \cite{meltzoff2013imitation}. Inspired by how intelligent creatures learn, IL tries to reproduce expert behaviors by utilizing expert policies or expert data \cite{Bain95,Ross11,piot2017brading}. IL has successfully solved many complex tasks that could be difficult for other methods, such as animation \cite{peng2018deepmimic}, robotic manipulation \cite{fang2019survey}, self-driving~\cite{codevilla2018end}, anomaly detection \cite{li2021automated}, etc. According to the imitation mechanism, IL can be divided into two categories \cite{hussein2017imitation}, behavioural cloning (BC) and inverse reinforcement learning (IRL).

BC regards the IL problem as a supervised learning (SL) problem \cite{Bain95}, aiming to obtain policies that can predict the correct actions (output) given states (input). Hence, BC learns in an off-line manner like SL, which indicates that BC does not demand interactions with environments. BC takes expert data as independent and identically distributed (IID). However, most IL problems are based on MDP, meaning that data are sequential and not independent and identically distributed (Non-IID) \cite{abbeel2005exploration}. The Non-IID property could lead to compounding errors and even failures of BC. To tackle the compounding error, Ross \emph{et al.}~\cite{Ross11} propose DAGGER, which alternatively samples from expert policies and agent policies. A basic difference between BC and DAGGER is that the former learns with only expert data, while the latter needs expert policies and on-line interactions.

IRL deals with the IL problem from the aspect of RL~\cite{abbeel2004apprenticeship}. There are two alternate steps in IRL: (1) learn a reward function that can assign high rewards to expert data and low rewards to agent's experiences; (2) conduct RL with the learned reward function in step (1) to improve the performance of agents. Abbeel and Ng \cite{abbeel2004apprenticeship} present apprenticeship learning, which employs linear functions to approximate the reward function. Maximum entropy IRL \cite{ziebart2008maximum} uses probabilistic models to obtain a reward function with less ambiguity based on maximum entropy and maintains performance guarantees. Instead of assuming linear rewards, GAIL \cite{ho2016generative} leverages DNNs to represent the reward function (also called the discriminator) and shows dramatic performance improvements. For more recent advances in IRL, please refer to \cite{torabi2018generative,ghasemipour2020divergence,zhang2020f,dadashi2021primal}.

\subsection{Quantum Machine Learning (QML)}
With remarkable progresses of quantum technologies in recent years, researchers have proposed to employ them in the domain of machine learning (ML), initiating a new research field---quantum machine learning (QML) \cite{biamonte2017quantum}. A fundamental motivation for QML is to speed up traditional ML algorithms that require huge computation with the potential quantum advantage. Quantum technologies have been successfully applied to the three main sub-domains of ML, \emph{i.e.}, supervised learning (SL) \cite{hastie2009overview}, unsupervised learning (USL)~\cite{hastie2009unsupervised}, and RL \cite{sutton1998introduction}. In SL, for example, quantum principal component analysis (QPCA) \cite{lloyd2014quantum}, quantum support vector machine (QSVM)~\cite{rebentrost2014quantum,ding2021quantum}, and quantum convolutional neural networks (QCNN)~\cite{oh2020tutorial} are proposed; in USL, clustering and \emph{k}-medians are accelerated with the quantum paradigm \cite{aimeur2013quantum,shrivastava2020classical}; in RL, there are \cite{dong2008quantum,saggio2021experimental,jerbi2021parametrized,skolik2021quantum}. However, quantum algorithms with theoretical speed-up guarantees are difficult to implement on current NISQ devices \cite{hsiao2022unentangled}. Limited by the NISQ era, one of the most promising methods that may achieve the potential quantum advantage is variational quantum circuits (VQCs) (also referred to as variational quantum algorithms, VQAs)~\cite{cerezo2021variational}. QML is still in its infancy and remains an open question that requires further efforts to prove and discover the quantum advantage, especially for VQA-based algorithms.

On the other hand, ML, which demonstrates super power in solving complex tasks such as Go and video games, also helps improve quantum technologies. Especially, among ML, RL has been applied to quantum control~\cite{bukov2018reinforcement}, qubit routing \cite{herbert2018using}, and circuit design. For example, Ostaszewski~\emph{et al.}~\cite{ostaszewski2021reinforcement} employ RL to optimize the structure of variational quantum eigensolvers (VQEs), which achieves state-of-the-art performance regarding the circuit depth.

Nonetheless, IL, a significant  branch of ML, has not been successfully enhanced with quantum technologies. Despite tremendous successes, IL usually constructs policies with DNNs and suffers from the high sample complexity, requiring huge computation resources when run on classical computers. Hence, we aim to empower IL with quantum advantage, which could improve the sample efficiency of IL and can pave the way for the future quantum era. To the best of our knowledge, we are the first to propose quantum imitation learning (QIL) and shed some light on this topic.

\section{Background}\label{sec:background}
We introduce some background knowledge at first, which is necessary for the presentation of our QIL algorithms.

\subsection{Imitation Learning}
\textbf{Markov Decision Process (MDP) \cite{puterman1994markov}.} An MDP can be described by a 5-tuple $(S,A,T,r,\gamma)$, where $S$ is the state space, $A$ represents the action space, $T=T(s'|s,a)$ depicts the environment dynamics by modeling the probability of state transitions over actions, $r:S \times A \rightarrow \mathbb{R}$ is the reward function, and $\gamma$ is a discount factor that $0\le \gamma \le 1$. The reward function is assumed to be bounded such that $|r(s,a)|\le R_{MAX}$, $\forall (s,a)\in S \times A$ \cite{haarnoja2018soft}. Assume the agent takes actions with a stochastic policy $\pi(a_t|s_t):S\times A\rightarrow[0,1]$ \cite{haarnoja2018soft}, where $t$ is the current timestep, then the expected discounted reward is defined as $J(\pi)=\mathbb{E}_{s_{0},a_{0},\cdots}\left[\sum_{t=0}^{\infty}\gamma^t r(s_t, a_t)\right]$, where $s_0\sim\rho_0(s_0)$, $a_t\sim\pi(a_t|s_t)$, $s_{t+1}\sim T(s_{t+1}|s_t,a_t)$, and the initial state $s_0$ is sampled from the probability distribution $\rho_0(s_0)$. RL algorithms aim to find the optimal policy $\pi^{*}(a_t|s_t)$ that can achieve the maximum episodic cumulative reward $J^*(\pi)$ via trial and error. \par

Compared to RL, in the IL setting, agents cannot receive rewards that are fed back from the environment. By contrast, agents are provided with some expert data or expert policies and demanded to reproduce expert behaviors \cite{hussein2017imitation}. We formalize the IL problem as follows.

\par
\textbf{Problem formulation of IL.}
We focus on IL from expert data rather than expert policies in this paper because the former is more common and practical~\cite{cheng2021guaranteed}. Given $N$ expert demonstrations $\mathcal{E}=\{(s_i,a_i)_{i \in N}\}$ sampled from an expert policy $\pi_E$, the optimizing objective for IL is defined \cite{osa2018algorithmic},
\begin{align}
    \pi^{*}=\arg \min D(q_{E}(s,a), p_{\pi}(s,a)),
\end{align}
where $q_{E}$ and $p_{\pi}$ are the distributions of the expert and agent, respectively, and $D$ is a distance measure such as mean squared error (MSE), Kullback–Leibler (KL) divergence \cite{kullback1951information}, Jensen–Shannon (JS) divergence \cite{lin1991divergence}, f-divergence \cite{renyi1961measures}, etc.

For subsequent presentations, we define the state and state-action occupancy measures as follows \cite{cheng2021guaranteed,xu2021error}
\begin{align}
    d_{\pi}(s)&=(1-\gamma)\sum_{t=0}^{\infty} \gamma^t P(s_t=s|\pi)\\
    \rho_{\pi}(s,a)&=(1-\gamma)\sum_{t=0}^{\infty} \gamma^t P(s_t=s, a_t=a|\pi).
\end{align}

\textbf{Behavioural Cloning (BC) \cite{Bain95,xu2021error}} BC regards IL as an SL problem and can be formulated as
\begin{align}\label{eq:ideabc}
\begin{split}
    \mathop{\min}\limits_{\pi} \, &\mathbb{E}_{s\sim d_{\pi_E}} [D_{KL}(\pi_E(\cdot|s)|\pi(\cdot|s))]:=\\
   &\mathbb{E}_{(s,a)\sim \rho_{\pi_E}} \left[\log \left( \frac{\pi_E(a|s)}{\pi(a|s)}\right)\right].
\end{split}
\end{align}
However, in practice, we usually have no access to the expert policy $\pi_E$ but only expert demonstrations $\mathcal{E}$. There are two commonly-used losses for BC with expert data, \emph{i.e.}, MSE loss and negative log-likelihood (NLL) loss. BC with NLL loss is presented as follows:
\begin{align}
    \pi^{*}=\arg \min \mathbb{E}_{(s^{E},a^{E})\sim \mathcal{E}}[-\log \pi(a^{E}|s^{E})].
\end{align}

\textbf{Generative Adversarial Imitation Learning (GAIL) \cite{ho2016generative}.} GAIL conducts IL by minimizing the divergence between expert trajectories and agent trajectories. GAIL works with a GAN-like structure \cite{goodfellow2014generative}, \emph{i.e.}, alternatively trains a discriminator and a generator (the agent's policy). The outputs of the discriminator measure the similarity of a state-action pair $(s,a)$ to the expert one, which are used as rewards to improve the generator. In particular, GAIL is formalized as follows:
\begin{align*}
    \begin{split}
        \mathop{\min}\limits_{\theta} \mathop{\max}\limits_{\omega} \mathop{\mathbb{E}}\limits_{s,a \sim \rho_\theta}&[\log D_\omega(s,a)]\\+&\mathop{\mathbb{E}}\limits_{s,a \sim \rho_{E}}[\log(1- D_\omega(s,a))] -\beta \mathcal{H}(\pi_{\theta}),
    \end{split}
\end{align*}
where $\rho_\theta$ and $\rho_{E}$ are the occupancy measures of the expert and agent, respectively, $D_\omega(s,a)$ is a discriminator, $\omega$ is the parameter of $D_\omega(s,a)$, $\mathcal{H}(\pi)=\mathbb{E}_{\pi}[-\log \pi(a|s)]$ is the entropy of policy $\pi$, and $\beta \ge 0$. The entropy term $\mathcal{H}(\pi)$ serves as a regularizer to stimulate exploration in GAIL.  

\subsection{Basics of Quantum Computing}

\subsubsection{Basics and Notations in Quantum Computing} 
We introduce some basic terminologies and notations of quantum computing employed in this work. Similar to the classical \emph{bit} which could be $0$ or $1$, a qubit \cite{de2019quantum} has two basis states $\ket{0} = (1,0)^\top$ and $\ket{1} = (0,1)^\top$. Being described as the quantum superposition, the general state of a qubit lies in the space spanned by basis states, \emph{i.e.}, $\ket{\psi} = \alpha \ket{0}+\beta \ket{1}$, where $\alpha,\beta \in \mathbb{C}$ and  $|\alpha|^2+|\beta|^2=1$. 
Thus, the state of a qubit could be represented in the space $\mathbb{C}^2$. The space for describing an $n$-qubit system is the tensor product of involved single qubit space, which has the dimension $M=2^n$. 
For example, the state generated by two qubits $\ket{\psi_1}=\alpha_1 \ket{0}+ \beta_1 \ket{1}$ and $\ket{\psi_2}=\alpha_2 \ket{0}+ \beta_2 \ket{1}$ is calculated as follows
\begin{align}\label{eq:tensorproduct}
\begin{split}
    \ket{\Psi}&=\ket{\psi_1} \otimes \ket{\psi_2}\\
    & =(\alpha_1 \ket{0}+ \beta_1 \ket{1}) \otimes (\alpha_2 \ket{0}+ \beta_2 \ket{1})\\
    & =\alpha_1 \alpha_2 \ket{00} + \alpha_1 \beta_2 \ket{01} + \beta_1 \alpha_2 \ket{10} + \beta_1 \beta_2 \ket{11}.
\end{split}
\end{align}
In general, an $n$-qubit quantum state can be represented by
\begin{align}
\begin{split}
    \ket{x} = &x_{1}\ket{\underbrace{0\cdot \cdot \cdot 0}_{n}}+x_{2}\ket{0\cdot \cdot \cdot1}+...+x_{2^n}\ket{1\cdot \cdot \cdot1}\\
    &s.t. \sum_{i=1}^{2^n} |x_{i}|^2=1 \;\text{and} \; x_{i} \in \mathbb{C}.
\end{split}
\end{align}
As presented in Eq. \eqref{eq:tensorproduct}, the 2-qubit state $\ket{\Psi}$ is produced with the tensor product of two single-qubit states $\ket{\psi_1}$ and $\ket{\psi_2}$. For instance, $\ket{01}=\ket{0} \otimes \ket{1}$. However, there exist 2-qubit states $\ket{\Psi}$ (for example, $\frac{\ket{00}+\ket{11}}{\sqrt{2}}$ and $\frac{\ket{01}+\ket{10}}{\sqrt{2}}$) that cannot be decoupled into the tensor product of two single-qubit states, which is defined as the quantum entanglement \cite{lockwood2020reinforcement}. For an un-entangled multi-qubit state $\ket{\Psi}$, every qubit channel will not be influenced by each other. On the contrary, for entangled ones, each qubit channel can affect the others such that we can control multiple qubits with only one operation. The exponential space size resulted from superposition and entanglement could provide the potential of quantum advantage \cite{preskill2012quantum, hirvensalo2013quantum} against classical counterparts.

Classical computing uses logic gates (such as AND, OR, and NOT) to perform basic logic operations on bits \cite{maini2007digital}. Similarly, quantum gates, based on Hamiltonian evolution, are defined to achieve logic operations on qubits \cite{divincenzo1998quantum}. Common quantum gates for 1-qubit system are Hadamard gate, Pauli gates $X$, $Y$, and $Z$, and parameterzied rotation gates where Pauli operators are served as Hamiltonians. The most fundamental gate for 2-qubit devices is CNOT, which can control the second qubit according to the first one,
\begin{align}
\begin{split}
    \text{CNOT} \ket{00}=\ket{00}, \text{CNOT} \ket{01}=\ket{01},\\
    \text{CNOT} \ket{10}=\ket{11}, \text{CNOT} \ket{11}=\ket{10}.\nonumber
\end{split}
\end{align}
The operation of quantum gates can be interpreted by matrix multiplications. For example, we denote the 2-qubit state $\ket{10}$ with a vector $(0,0,1,0)^\top$. Then, the CNOT gate is described by a $4\times 4$ matrix, where $4=2^2$,
\begin{align}
    \text{CNOT} =  \begin{bmatrix} 1 & 0 & 0 & 0 \\ 0 & 1 & 0 & 0 \\ 0 & 0 & 0 & 1 \\ 0 & 0 & 1 & 0 \end{bmatrix}. \nonumber
\end{align}
Hence, we obtain $\text{CNOT} \ket{10}=\ket{11}$. In particular, it is clear that $\text{CNOT} \cdot \text{CNOT}^\dagger=I$. In quantum computing, an operation $U$ for the $n$-qubit system could be represented by a $2^n \times 2^n$ unitary matrix, \emph{i.e.}, $UU^\dagger=U^\dagger U=I$ \cite{brylinski2002universal}. Resembling classical logic circuits, we can construct quantum circuits with basic quantum gates to perform operations on multi-qubit quantum systems.

\subsubsection{Variational Quantum Circuits}
Quantum circuits, constructed with a sequence of quantum gates, are computational routines to perform quantum operations \cite{xiang2013hybrid}. There are no adjustable parameters in traditional quantum circuits. In contrast, DNNs, which can receive input data and generate non-trivial outputs, are built with learnable parameters \cite{yosinski2014transferable}. The corresponding structure in quantum technology to replace DNNs is variational quantum circuits (VQCs) \cite{benedetti2019parameterized}. VQCs are quantum circuits controlled with learnable parameters $\mu$, which can accept input data with state preparation and generate outputs with quantum measurement. It has been proved that VQCs have the ability to universally approximate functions like DNNs \cite{biamonte2021universal}. Furthermore, in the NISQ era, VQCs have demonstrated potentials to outperform DNNs regarding model size and mitigating overfitting \cite{chen2021end}.

A VQC for an $n$-qubit system is formalized as follows
\begin{align}
    f(s,\mu)=\bra{0^{\otimes n}} U^{\dagger}(s,\mu) O U(s,\mu)\ket{0^{\otimes n}},\label{f_formulation}
\end{align}
where $s$ is the input state, $O$ is a quantum observable, and $U(s,\mu)$ is a unitary operation specified with state $s$ and parameter $\mu$. To obtain outputs of VQCs, there are three procedures \cite{benedetti2019parameterized,lockwood2020reinforcement}: (1) prepare states via encoding to convert classical data to quantum data; (2) entangle qubits with a variational circuit, whose parameters are controllable; (3) to acquire classical information, measure quantum states with quantum gates such as Pauli Z gate. After measurement in procedure (3), we can use the readout information to construct loss functions as in traditional ML. One remaining question is how to update parameters in VQCs. For the case that the parameter $\mu$ is employed as the phase of quantum gates, we use the parameter-shift rule \cite{benedetti2019parameterized} to calculate gradients. For example, we can calculate the partial derivative with respect to the parameter vector $\mu_j$ by
\begin{align}
    \frac{\partial f(s,\mu)}{\partial  \mu_j}=  f(s,\mu+\frac{\pi}{4} e_j )-f(s,\mu-\frac{\pi}{4} e_j ),
\end{align}
where the unitary $U(s,\mu)$ in Eq.~\eqref{f_formulation} contains the trainable gate $e^{-i \mu_j H_j}$, and the hermitian matrix $H_j$ serves as the Hamiltonian of the gate.

\section{Methods}\label{sec:method}
In this section, we introduce the proposed QIL algorithms, Q-BC and Q-GAIL, in detail. First, we construct modified VQCs that can replace DNNs with special considerations of the characteristics of IL; second, the constructed VQCs are employed in Q-BC to serve as the policy network, and the error bound of Q-BC is derived; third, we apply VQCs to IRL and present the Q-GAIL algorithm.

\subsection{Variational Quantum Circuits for Imitation Learning}
The ultimate objective of IL is to learn a policy $\pi(a|s)$ that can generate right actions to behave as similarly as experts. In other words, IL learns a function mapping from states to actions. In QIL, we aim to employ VQCs $f(s,\mu)$ to build policies, making it possible to obtain quantum advantage. Besides, policies using VQCs could be run on NISQ devices, which could utilize the power of quantum computing and improve the sample efficiency. To adapt VQCs for IL, we modify traditional VQCs from three aspects: (1) encode traditional data for direct delivery to VQCs (Subsection \ref{subsec:dataencoding}); (2) stack multiple layers of quantum gates with data re-uploading to improve the representation capability of VQCs (Subsection \ref{subsec:datareuploading}); (3) modify outputs of VQCs to achieve better exploration (Subsection \ref{subsec:datareadout}).

\subsubsection{Data Encoding} \label{subsec:dataencoding}
Unlike classical computers that can directly deal with data sampled from environments, the first encountered problem to conduct QIL is how to encode data for the ease of quantum computing. A straightforward way is to convert traditional data into binary and consequently represent them with qubits \cite{weigold2021expanding}, which is dubbed as basis encoding. Basis encoding is convenient for applications in discrete state space. However, it remains challenging for basis encoding to represent data in continuous environments. For example, a float-point number typically requires 4 bytes, \emph{i.e.}, 32 bits, to store in classical computers. As a result, basis encoding needs 32 qubits to encode a float-point number, which is inefficient and not suitable for the NISQ era where the number of qubits is limited. To tackle the problem, there are other data encoding schemes in quantum computing include qubit encoding \cite{schuld2019quantum} (also known as angle encoding), amplitude encoding \cite{prakash2014quantum}, etc.

In this work, we exploit qubit encoding for our VQCs. Qubit encoding is efficient regarding operations because it only requires one single parallel operation no matter how many qubits there are \cite{weigold2021expanding}. Moreover, it demands less qubits to encode data compared to basis encoding. For a given classical array $X=\{x_1,\dots,x_n\}$, we first normalize each item $\{ x_i| 0 \le i \le n\}$ to the range $[-\pi,\pi]$. Then, $X$ can be encoded via qubit encoding as follows
\begin{align}
    \ket{\Psi(X)}=\left(\begin{array}{c} \cos x_1 \\ \sin x_1 \\ \end{array} \right) \otimes \dots \otimes \left(\begin{array}{c} \cos x_n \\ \sin x_n \\ \end{array} \right).
\end{align}

\subsubsection{Data Re-uploading}\label{subsec:datareuploading}
In the above subsection, we introduce how to encode classical data such that it is possible to upload them to VQCs. In traditional IL, we stack multiple layers of neurons to improve the representation ability of DNNs~\cite{nguyen2020wide}. Besides, input data can be copied and processed multiple times by different neurons at a single layer. However, data cannot be accessed anymore once they have been uploaded to VQCs. After uploading, the information of original data is missing, and VQCs are only available to the rotations of qubits. A single rotation, which encodes classical data, is merely able to model simple sine functions \cite{skolik2021quantum} and is not enough to express complex pattern separations of initial data \cite{perez2020data}. Inspired by DNNs, Perez \emph{et al.} \cite{perez2020data} propose to re-upload data several times and sequentially stack them, aiming to improve the expressivity of VQCs. Concretely, in VQCs, we upload original data via one-qubit rotations and replicate the variational part that contains parameters several times. This data re-uploading structure bears some resemblance to the construction of DNNs because VQCs grow deeper with re-uploading. Furthermore, data re-uploading enables VQCs to universally approximate non-trivial functions, which dramatically improve the performance of classification \cite{perez2020data,perez2021one}.

In this paper, we employ data re-uploading to build imitator policies. Although data are uploaded several times, this re-uploading framework helps build circuits with fewer qubits. To further enhance the expressivity of QNNs \cite{jerbi2021parametrized,skolik2021quantum}, we adopt a scaling parameter $\lambda$. The scaling parameter $\lambda$ is applied to input data to dynamically adjust its range. For an array $X=\{x_1,\dots,x_n\}$, it is processed with parameter $\lambda=\{\lambda_1,\dots,\lambda_n\}$
\begin{align}
    x_i = \lambda_i \times x_i.
\end{align}
The parameter $\lambda$ can be dynamically updated during training. Up to now, we denote the trainable parameters in VQCs as $\mu=(\lambda, \phi)$, $\lambda$ is the scaling parameter while $\phi$ is the parameter in variational parts.

\subsubsection{Data Readout}\label{subsec:datareadout}
With data encoding and re-uploading, the information contained within agent's states can be processed in the quantum system. After processing, we need to read out the control signal from the quantum system to generate actions and apply these actions to the environment. The only way to convert quantum information into classical counterpart is quantum measurement \cite{de2019quantum}. We use $O$ to represent a quantum observable, which is related to the measurement operation. For example, Pauli gates, such as $Z$ gates, are commonly-used quantum observables to read out quantum outputs. The output of the observable is constrained within $[-1,1]$. Given an observable $O_a$, whose subscript denotes action $a$, a policy can be built with a VQC   
\begin{align}
\label{eq:rawvqc}
    \pi_{\theta} (a|s) = \bra{0^{\otimes n}} U^{\dagger}(s,\mu) O_a U(s,\mu)\ket{0^{\otimes n}}.
\end{align}
The VQC in Eq. \eqref{eq:rawvqc} is widely adopted in SL tasks, whereas the policy based on Eq. \eqref{eq:rawvqc} is poor at exploration for discrete-action tasks, which is vital in IL. To approach this issue, a softmax layer is added after the observable. This kind of VQC is coined as softmax-VQC \cite{jerbi2021parametrized}, which is presented below, 
\begin{align}
    \pi_{\theta} (a|s) = \frac{e^{\beta \bra{0^{\otimes n}} U^{\dagger}(s,\mu) O_a U(s,\mu)\ket{0^{\otimes n}}}}{\sum_{a'}e^{\beta \bra{0^{\otimes n}} U^{\dagger}(s,\mu) O_{a'} U(s,\mu)\ket{0^{\otimes n}}}},
\end{align}
where $O_a=\sum_i \nu_{a,i} H_{a,i}$,  $\nu_{a,i}$ is a learnable scaling parameter to further enhance the representation ability, $H_{a,i}$ represents a Hermitian operator related to action $a$, and $\beta \in \mathbb{R} \;(\beta > 0)$ is the inverse temperature to control the policy greediness. The trainable parameters $\theta=(\lambda, \phi, \nu)$ for a softmax-VQC can be divided into three parts, $\lambda$, $\phi$, and $\nu$, where $\theta=(\mu, \nu)$,  $\mu=(\phi,\lambda)$, and $\nu=\{\nu_{a,i}\}$. 

In summary, to construct policies with VQCs in IL, we first convert classical states into quantum states via qubit encoding; then the encoded states are multiplied by scaling parameters and re-uploaded several times; finally, quantum observables are used to read out control signals from a quantum system, followed by a softmax layer to stimulate exploration. The representation capability of softmax-VQC is significantly enhanced with scaling parameters and the data re-uploading scheme to accommodate IL policies. 

When the action of the environment becomes continuous, the softmax-VQC is not feasible to represent policies. Hence, we remove the softmax layer of the softmax-VQC and propose Gaussian-VQC as follows,
\begin{align*}
    \pi_{\theta} (a|s) \sim N( \bra{0^{\otimes n}} U^{\dagger}(s,\mu) O_a U(s,\mu)\ket{0^{\otimes n}}, \sigma^2),
\end{align*}
where $N(\cdot,\cdot)$ is the normal distribution, and $\sigma$ is a learnable parameter and represents the standard deviation. In Gaussian-VQC, the trainable parameter is $\theta=(\lambda, \phi, \nu, \sigma)$.

\subsection{Quantum Behavioural Cloning (Q-BC)}
Using the policy constructed in the above subsection, we design the QIL algorithm based on behavioural cloning, \emph{i.e.}, Q-BC, which is presented in Algorithm \ref{algo:Q-BC}. Compared to traditional BC, Q-BC employs a policy that is built with softmax-VQCs. Hence, to better encode data for softmax-VQCs, both states from the expert and agent should be normalized with state bounds before uploading to quantum systems. In Q-BC, there are only expert states that require normalization. For a given state bound $s_b$, we can normalize the state by $s = s/s_b$.
\begin{algorithm}[tb] 
  \caption{Quantum Behavioural Cloning (Q-BC)}
\begin{algorithmic}
  \STATE {\bfseries Input:} Expert demonstrations $\mathcal{E}$, iteration number $m$, learning rate $\xi_\theta$, state bound $s_b$, and minibatch size $N$.
  \STATE {\bfseries Parameter:} Policy $\pi_{\theta}$, where $\theta=(\lambda, \phi, \nu)$.
      \FOR{$i=1$ {\bfseries to} $m$}
          \STATE Sample $N$ transitions from expert demonstrations $$\{(s^{E},a^{E})\}_{1}^{N} \sim \mathcal{E}$$
          \STATE Normalize sampled expert states $s^{E}=s^{E}/s_b$
          \STATE Calculate gradients of $\pi_{\theta}$ $$\nabla_{\theta} \hat L=- \frac{1}{N} \nabla_{\theta} \log \pi_\theta(a^{E}|s^{E})$$
          \STATE Update policy parameters $$\theta := \theta - \xi_\theta \nabla_{\theta} \hat L$$
      \ENDFOR
\end{algorithmic}
\label{algo:Q-BC}
\end{algorithm}
After normalization, each dimension of $s$ is constrained within $[-\pi,\pi]$ radian, which can be subsequently encoded using qubit encoding. With normalized states uploaded to the quantum system, the policy can infer the probability of choosing an action, which is further optimized with an NLL loss. The loss function of Q-BC is presented as follows,
\begin{align}
    L(\theta) = \mathbb{E}_{(s^{E}, a^{E})\sim \mathcal{E}}[-\log \pi_{\theta}(a^{E}| s^{E})]. 
\end{align}
Consequently, the derivative $\nabla_{\theta} L$ is
\begin{align}
    \nabla_{\theta} L &= \mathbb{E}_{(s^{E},a^{E})\sim \mathcal{E}}\nabla_{\theta}[-\log \pi_{\theta}(a^{E}| s^{E})].
\end{align}
The gradient of the log-probability of a policy $\pi_{\theta}$ can be calculated with the following lemma.
\begin{lemma}\label{lemma:gradoflogprob}
 \cite{jerbi2021parametrized} For a policy $\pi_{\theta}$ constructed with softmax-VQCs, the gradient of the log-probability can be obtained,
 \begin{align*}
 \begin{split}
    \nabla_{\theta} \log \pi_{\theta}(a^{E}| s^{E}) &= \frac{\nabla_{\theta} \pi_{\theta}(a^{E}| s^{E})}{\pi_{\theta}(a^{E}| s^{E})}\\
    &=\beta\left(\nabla_{\theta} \langle O_a\rangle_{s,\theta}-\sum_{a'}\pi_{\theta} (a'|s) \langle O_{a'}\rangle_{s,\theta}\right),
 \end{split}
 \end{align*}
 where $\langle O_a\rangle_{s,\theta}$ denotes $\bra{0^{\otimes n}} U^{\dagger}(s,\mu) O_a U(s,\mu)\ket{0^{\otimes n}}$. 
\end{lemma}
Utilizing Lemma \ref{lemma:gradoflogprob}, we can sample expert data and obtain the estimated gradient $\nabla_{\theta} \hat L$ for updating the imitator policy in Q-BC. The parameter of the policy is updated via an incremental manner
\begin{align}
    \theta_{k+1} = \theta_k - \xi_\theta \nabla_{\theta} \hat L, 
\end{align}
where $\xi_\theta$ is the learning rate, and $k$ is the current iteration step.

With quantum policies in the NISQ era, it is hard to accurately measure quantum states, which could bring errors in the optimization and affect the IL performance. Hence, we would like to analyze the impact of quantum measurement errors on the IL performance. Consequently, an assumption is made on the measurement error.

\begin{assumption}\label{assumption:quantummeasureerror} \cite{jerbi2021parametrized}
We have efficient approximations $\langle \widetilde O_a\rangle_{s,\theta}$ measured from the unknown true expectation values $\langle O_a\rangle_{s,\theta}$, which can be described by a most additive measurement error $\epsilon$ such that
\begin{align}
    \left| \langle \widetilde O_a\rangle_{s,\theta} - \langle O_a\rangle_{s,\theta} \right| \le \epsilon, \forall a \in A.
\end{align}
\end{assumption}
Assumption \ref{assumption:quantummeasureerror} assumes the measurement error is bounded, which is reasonable in most realistic quantum systems. Besides, we can independently repeat executions and measurements with $\mathcal{O}(\epsilon^{-2})$ times on a quantum computer to reduce the error. As a result, to distinguish with the true policy $\pi_{\theta}$, we denote the approximated policy as $\widetilde \pi_{\theta}$,
\begin{align}
    \pi_{\theta}= \frac{e^{\beta \langle O_{a}\rangle_{s,\theta} }}{\sum_{a'}e^{\beta \langle O_{a'}\rangle_{s,\theta}}}, \widetilde \pi_{\theta}= \frac{e^{\beta \langle \widetilde O_{a}\rangle_{s,\theta} }}{\sum_{a'}e^{\beta \langle \widetilde O_{a'}\rangle_{s,\theta}}}. 
\end{align}
After optimizing BC loss under the infinite sample situation, the difference between the agent approximated policy $\widetilde \pi_{\theta}$ and expert policy $\pi_E$ is depicted with Assumption \ref{assumption:KLdivergencebetweenlearnerandexpert}. 

\begin{assumption}\label{assumption:KLdivergencebetweenlearnerandexpert}
\cite{xu2021error} We can achieve a $\delta$ error by optimizing the BC loss in Eq. \eqref{eq:ideabc} such that
\begin{align}
    \mathbb{E}_{s \sim d_{\pi_E}}[D_{KL}(\pi_E(\cdot |s),\widetilde \pi_\theta(\cdot |s))] \le \delta.
\end{align}
\end{assumption}
This assumption indicates that the output policy $\widetilde \pi_\theta(\cdot |s)$ is close to the expert policy $\pi_E(\cdot |s)$, which is a common routine in IL \cite{Ross10,rajaraman2020toward}. Based on Assumptions \ref{assumption:quantummeasureerror} and \ref{assumption:KLdivergencebetweenlearnerandexpert}, we derive an error bound for the algorithm Q-BC, formalized in Theorem~\ref{theorem:bcbound}. 
\begin{theorem}\label{theorem:bcbound}
Given an expert policy $\pi_E$ and the true learner policy $\pi_\theta$ generated by Q-BC, suppose Assumptions 1 and 2 hold, the error bound is
\begin{align}
    |J(\pi_E)-J(\pi_\theta)|\leq \frac{2R_{MAX}}{(1-\gamma)^2}(|\sinh(2\beta \epsilon )| + \sqrt{2} \sqrt{\delta}),
\end{align}
where $\sinh$ is the hyperbolic sine function. 
\end{theorem}
The proof is deferred to the Appendix. From Theorem~\ref{theorem:bcbound}, we can see that the error bound grows exponentially regarding the quantum measurement error $\epsilon$ and grows linearly with respect to the square root of the optimization error $\delta$. Besides, Theorem \ref{theorem:bcbound} indicates that Q-BC achieves an error that is quadratic with $\frac{1}{1-\gamma}$, which is the effective planning horizon~\cite{xu2021error}. We also notice that the error bound will shrink with a smaller $\beta$.

\subsection{Quantum Generative Adversarial Imitation Learning (Q-GAIL)}
Here, we introduce another QIL algorithm, Q-GAIL, which is presented in Algorithm \ref{algo:Q-GAIL}. In Q-GAIL, there are two main blocks: one block is the discriminator, while the remaining is the generator. The discriminator outputs a scalar $D_\omega(s,a) \in [0,1]$, which judges the quality of state-action pairs sampled from the generator. In the IL setting, there are no rewards provided by the environment, and the discriminator is employed to deal with the lack of a reward function, whose outputs are referred to as virtual rewards. The generator contains an RL algorithm to optimize a policy with virtual rewards from the discriminator. During training, the discriminator and generator are updated alternately.

In this paper, we regard the discriminator as a part of the environment and implement it with traditional DNNs as in~\cite{huang2021experimental}, which is trained with the loss
\begin{align}
  \mathop{\max}\limits_{\omega} \mathop{\mathbb{E}}\limits_{s,a \sim \rho_\theta}[\log D_\omega(s,a)]+\mathop{\mathbb{E}}\limits_{s,a \sim \rho_{E}}[\log(1- D_\omega(s,a))].
\end{align}
It is also feasible to build a VQC-based discriminator for Q-GAIL, which has been employed in \cite{lloyd2018quantum} and is validated in Section \ref{sec:experiments}. In contrast to Q-BC, there are expert states and agent states in Q-GAIL, both of which need to be normalized. Furthermore, to alleviate the instability caused by the adversarial training scheme, we adopt spectral normalization for the discriminator \cite{miyato2018spectral, cheng2021guaranteed}. Denote a weight matrix of $\omega$ as $W$, its spectral norm can be obtained
\begin{align}
    \sigma(W):=\mathop{\max}\limits_{\Arrowvert h\Arrowvert_2 \le 1}\Arrowvert Wh\Arrowvert_2.
\end{align}
After each optimization step, we spectrally normalize weight matrices of $\omega$ to limit the Lipschitz constant of $D_\omega$ by
\begin{align}\label{eq:spectralnorm}
    W_{SN}:=W/\sigma(W), W \in \omega.
\end{align}
This technique is vital for the stabilization and success of Q-GAIL, which is ablated in Section \ref{sec:experiments}.

On the contrary, the policy is implemented with softmax-VQCs to utilize quantum advantage. Besides, we use \textbf{REINFORCE} \cite{sutton1998introduction}, one of the most fundamental algorithms in RL, to learn from virtual rewards. It should be noted that \textbf{REINFORCE} in Q-GAIL can be replaced by other RL algorithms such as TRPO \cite{schulman2015trust} or PPO \cite{schulman2017proximal}  as we demonstrated in the experiments.  In particular, we rewrite the objective function $J^{Q}(\pi_\theta)$ for the quantum generator as
\begin{align}
\begin{split}
J^{Q}(\pi_\theta)=\mathbb{E}_{s_{0},a_{0},\cdots}\left[\sum_{t=0}^{\infty}\gamma^t \bar r(s_t, a_t)\right]
    \!\!=\mathbb{E}_{\tau \sim \pi_\theta}\left[\bar R(\tau)\right],
\end{split}
\end{align}
where $\tau=(s_0,a_0,\cdots)$ represents a trajectory sampled from the policy $\pi_\theta$, $\bar r$ is a virtual reward function, and $\bar R(\cdot) = \sum_{t=0}^{\infty}\gamma^t \bar r(s_t,a_t)$ calculates the cumulative virtual rewards of a trajectory. In Q-GAIL, the virtual reward can be obtained with different reward functions. For example, the most commonly-used reward function is $\bar r(s,a) = - \log D_\omega(s,a)$. Other reward definitions could be $\bar r(s,a) = \log (1- D_\omega(s,a))$ or $- (\log D_\omega(s,a)-\log (1- D_\omega(s,a)))$ \cite{hussenot2021hyperparameter}. For different applications, we sweep the above reward functions to select the most suitable one to achieve satisfactory IL performance. 

With virtual rewards from the discriminator, we can obtain the gradient of the optimization objective $J(\pi_\theta)$ with policy gradient theorem \cite{sutton1998introduction,SpinningUp2018} as follows
\begin{align}\label{eq:policygradienttheorem}
\begin{split}
    \nabla_\theta J^{Q}(\pi_\theta)&\!\!=\!\nabla_\theta \mathbb{E}_{\tau \sim \pi_\theta}\left[\bar R(\tau)\right]\\
    &\!\!=\!\mathbb{E}_{\tau \sim \pi_\theta}\left[\sum_{t=0}^{\infty} \nabla_\theta \log \frac{e^{\beta \langle O_{a}\rangle_{s,\theta} }}{\sum_{a'}e^{\beta \langle O_{a'}\rangle_{s,\theta}}} \bar R(\tau)\right]\!\!.
\end{split}
\end{align}
The calculation of $\nabla_\theta \log \pi_{\theta}(a | s)$ is presented in Lemma \ref{lemma:gradoflogprob}, and $\bar R(\tau)$ has no dependence on $\theta$. It is difficult to accurately calculate the policy gradient because an expectation over a distribution is required. Instead, we use samples from the policy $\pi_\theta$ to estimate the gradient. Given a set of trajectories $\mathcal{D}=\{ \tau_i \}_{i=0}^{N}$, the estimation of policy gradient can be obtained
\begin{align}
    \hat \nabla_\theta J^{Q}(\pi_\theta)=\frac{1}{\left | \mathcal{D} \right |} \sum_{\tau \in \mathcal{D}} \sum_{t=0}^{\infty} \nabla_\theta \log \frac{e^{\beta \langle O_{a}\rangle_{s,\theta} }}{\sum_{a'}e^{\beta \langle O_{a'}\rangle_{s,\theta}}}  \bar R(\tau).
\end{align}

To reduce variance in the RL algorithm \textbf{REINFORCE}, we replace $\bar R(\tau)$ with a more stable version as in \cite{SpinningUp2018},
\begin{align}\label{eq:gradientofpolicy}
\begin{split}
    \hat \nabla_\theta &J^{Q}(\pi_\theta)=\\
    &\frac{1}{\left | \mathcal{D} \right |} \sum_{\tau \in \mathcal{D}} \sum_{t=0}^{\infty} \nabla_\theta \log \frac{e^{\beta \langle O_{a}\rangle_{s,\theta} }}{\sum_{a'}e^{\beta \langle O_{a'}\rangle_{s,\theta}}} (\bar R_t (\tau) - b(s_t)), 
\end{split}
\end{align}
where $\bar R_t (\tau) := \sum_{t'=t}^{\infty} \gamma^{t'-t} \bar r (s_{t'}, a_{t'})$ is the discounted cumulative rewards for a trajectory after timestep $t$, and $b(s_t)$ is called a \textit{baseline}, which could be any function that does not rely on the action $a$. We select the value function $V_{\psi}^{\pi_\theta}(s)$ as the \textit{baseline}, which outputs a scalar of the predicted value function. Besides, the value function $V_{\psi}^{\pi_\theta}(s)$ is linear with respect to time-varying features as in \cite{duan2016benchmarking}. 
\begin{algorithm}[tb]
  \caption{Quantum Generative Adversarial Imitation Learning (Q-GAIL)}
\begin{algorithmic}
  \STATE {\bfseries Input:} Expert demonstrations $\mathcal{E}$, iteration number $m$, learning rate $\xi_\theta$, $\xi_\omega$, and $\xi_\psi$, inverse temperature $\beta$, state bound $s_b$, and trajectory size $N$.
  \STATE {\bfseries Parameter:} Policy $\pi_{\theta}$, discriminator $D_{w}$, \textit{baseline} $V_{\psi}^{\pi_\theta}(s)$.
      \FOR{$i=1$ {\bfseries to} $m$}
          \STATE Sample $N$ agent trajectories $\mathcal{D}=\{\tau\}_0^N$ under policy $\pi_{\theta}$ and denote the state-action pairs as $\{(s,a)\}_{1}^{n}$
          \STATE Sample $n$ expert transitions $\{(s^{E},a^{E})\}_{1}^{n} \sim \mathcal{E}$
          \STATE Normalize expert states and agent states
          $$x=x/s_b, x=s \text{ or } s^E$$
          \STATE Obtain rewards with the discriminator
          $$\bar r(s,a) = \log(1- D_\omega(s,a))$$
          \STATE Calculate gradients of $D_\omega$
          $$\hat \nabla_{\omega}J(D_\omega)=\frac{1}{m} \nabla_{\omega}(\log D_\omega(s,a)+ \log(1- D_\omega(s^E,a^E)))$$
          \STATE Calculate gradients of $\pi_\theta$ and $V_{\psi}^{\pi_\theta}(s)$, $\hat \nabla_\theta J^{Q}(\pi_\theta)$ and $\hat \nabla_\psi J(V_{\psi}^{\pi_\theta}(s))$, using Eqs. \eqref{eq:gradientofpolicy} and \eqref{eq:gradientofbaseline}
          \STATE Incrementally update parameters
          \begin{align}\notag
              \theta &= \theta + \xi_\theta \hat \nabla_\theta J(\pi_\theta) \\ \notag
              \omega &= \omega + \xi_\omega  \hat \nabla_{\omega}J(D_\omega) \\ \notag
              \psi &=  \psi - \xi_\psi \hat \nabla_\psi J(V_{\psi}^{\pi_\theta}(s))
          \end{align}
          \STATE Normalize $\omega$ with Eq. \eqref{eq:spectralnorm}
      \ENDFOR
\end{algorithmic}
\label{algo:Q-GAIL}
\end{algorithm}

During training, $V_{\psi}^{\pi_\theta}(s)$ is updated to fit the expected discounted return by minimizing the following loss, 
\begin{align}
     J(V_{\psi}^{\pi_\theta}(s))= \mathbb{\hat E} [(\sum_{t'=t}^{\infty}\gamma^{t'-t} \bar r(s_{t'},a_{t'})-V_{\psi}^{\pi_\theta}(s_t))^2]. 
\end{align}

\begin{figure*}[!htbp]
  \centering
      \subfigure[CartPole-v1]{
         \includegraphics[width=0.15\textwidth]{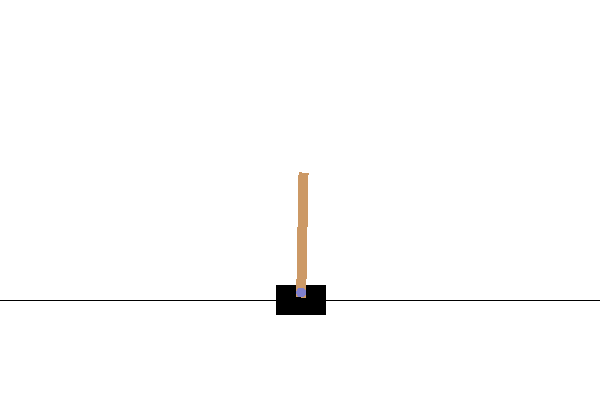}
         \label{fig:screenshotofcartpole}
      }
      \subfigure[Acrobot-v1]{
         \includegraphics[width=0.15\textwidth]{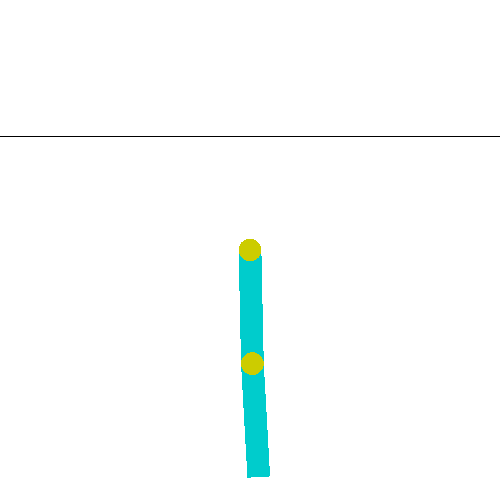}
      }
      \subfigure[MountainCar-v0]{
         \includegraphics[width=0.15\textwidth]{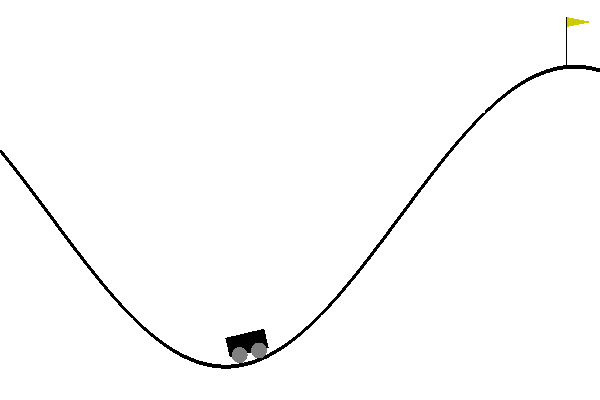}
      }
      \subfigure[InvertedPendulum-v2]{
         \includegraphics[width=0.15\textwidth]{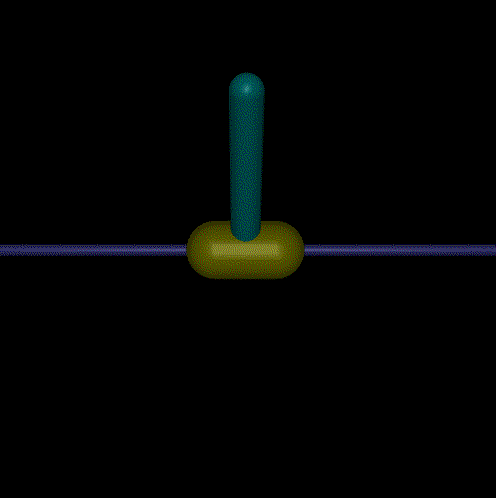}
      }
  \caption{Screenshots of employed OpenAI Gym environments.}
  \label{fig:screenshotofenv}
\end{figure*}

\begin{table*}[!htbp]
\centering
    \caption{Specifications of the OpenAI Gym tasks.}
\label{table:gymspecfication}
\centering
\begin{tabular}{l c c c c}
  
  \hline
  Environment     & State Dimension & State Range   & Feasible Actions & Max-Step\\
  \hline
  CartPole-v1 & 4 & $[-4.8, -\infty, -0.418, -\infty]\text{-}[4.8, +\infty, 0.418, +\infty]$ & 0/1  & 500\\
  Acrobot-v1 & 6 & $[-1, -1, -1, -1, -12.57, -28.27]\text{-}[1, 1, 1, 1, 12.57, 28.27]$ & -1/0/+1  & 500\\
  MountainCar-v0 & 2 & $[-1.2 , -0.07]\text{-}[0.6 , 0.07]$ & -1/0/+1  & 200\\
  InvertedPendulum-v2 & 4 & $[-\infty, -\infty, -\infty, -\infty]\text{-}[+\infty, +\infty, +\infty, +\infty]$ & [-3,3]  & 1000\\
  \hline
\end{tabular}
\end{table*}

The gradients with respect to parameters of $V_{\psi}^{\pi_\theta}(s)$ are also estimated with samples 
\begin{align}\label{eq:gradientofbaseline}
\begin{split}
     \hat \nabla_\psi &J(V_{\psi}^{\pi_\theta}(s))\\
     &\!\!=\!\!\frac{1}{\left | \mathcal{D} \right |} \sum_{\tau \in \mathcal{D}}\sum_{t=0}^{\infty} \nabla_\psi(\sum_{t'=t}^{\infty} \gamma^{t'-t} \bar r (s_{t'}, a_{t'}) - V_{\psi}^{\pi_\theta}(s_t))^2.
\end{split}
\end{align}
In summary, there are three blocks in Q-GAIL: a discriminator $D_\omega$, a policy $\pi_\theta$, and a \textit{baseline} $V_{\psi}^{\pi_\theta}(s)$. The discriminator serves as a virtual reward function, while the policy and the \textit{baseline} together are used as the generator. We have derived how to update the parameter of each block, and the complete algorithm of Q-GAIL is presented in Algorithm~\ref{algo:Q-GAIL}.

\section{Experiments}\label{sec:experiments}
In this section, we present the experimental results of our QIL algorithms, Q-BC and Q-GAIL, on OpenAI Gym tasks~\cite{brockman2016openai}. Three continuous-state and discrete-action environments (CartPole-v1, Acrobot-v1, and MountainCar-v0) are selected to demonstrate the performance of Q-BC and Q-GAIL. We also validate Q-BC and Q-GAIL with a continuous-action task, \emph{i.e.}, InvertedPendulum-v2. Furthermore, we conduct ablation studies to investigate what factors may affect the IL performance. Three important questions are focused:
\begin{itemize}
    \item Whether could Q-BC and Q-GAIL successfully learn from expert data and achieve expert-level performance?
    \item What are the impacts of the ingredients in Q-BC and Q-GAIL on the IL performance (including the number of expert data, the structure of softmax-VQCs, etc)?  
    \item Do QIL algorithms outperform the classical counterparts?
\end{itemize}

\subsection{Environments}
We adopt three OpenAI classic control tasks \cite{brockman2016openai} (\emph{i.e.,} CartPole-v1, Acrobot-v1, and MountainCar-v0) and one Mujoco task (InvertedPendulum-v2) for our experiments, which are presented in Fig. \ref{fig:screenshotofenv}. We take CartPole-v1 (Fig. \ref{fig:screenshotofcartpole}) as an example to introduce the employed environments. For more details about the other environments, please refer to~\cite{brockman2016openai}. In CartPole-v1, the state of the agent is continuous while the action is discrete. The state dimension of CartPole-v1 is 4, where the first and second dimensions represent the position and velocity of the cart, respectively, while the third and fourth dimensions are the angle and angle velocity of the pole. The action is discrete and could be 0 or 1, 0 for pushing the cart to the left and 1 for pushing to the right. The aim of control for CartPole-v1 is to keep the pole upright without falling over during an episode. The reward would be +1 for every timestep if an episode does not end, and the maximum length of an episode is 500. Hence, the optimal return of one episode is 500. However, during IL, we have no access to reward feed-backs. In contrast, agents are provided with expert data to enable imitating expert behaviors. The brief specifications of used environments are listed in Table~\ref{table:gymspecfication}. Note that the reward function in MountainCar-v0 is modified as in~\cite{duan2016benchmarking,jerbi2021parametrized}.

\begin{table}[!h]
\centering
    \caption{Performance of expert data.}
\label{table:expertperformance}
\centering
\begin{tabular}{l c c}
  \hline
  Environment     & Trajectories & Performance \\
  \hline
  CartPole-v1 & 200 & 494.0$\pm$21.2 \\
  Acrobot-v1 & 200 & -97.4$\pm$ 12.7\\
  MountainCar-v0 & 200 & -79.8 $\pm$ 6.3\\
  InvertedPendulum-v2 & 100 & 1000 $\pm$ 0.0\\
  \hline
\end{tabular}
\end{table}

\subsection{Expert Data}
In real-world applications, expert data are often sampled from experts like human beings with specific devices \cite{pan2020imitation}. For example, in robot locomotion, expert data for motions are collected with markers and high speed infrared cameras~\cite{de2009guide}. To conduct experiments of QIL, we first introduce how to obtain the expert data for classical control tasks. There are two steps to generate expert data: 1) train an expert policy with an RL algorithm; 2) execute the expert policy in the environment and record expert transition tuples including states and actions. Specifically, the RL algorithm in \cite{jerbi2021parametrized} (implemented in TensorFlow Quantum \cite{broughton2020tensorflow}) is employed to train expert policies for the classic control tasks. The maximum episode length is $500$ or $200$ for these three environments, and an episode of states and actions is dubbed as a trajectory. We sample trajectories for each task, and expert performance is presented in Table \ref{table:expertperformance}. As for the continuous-action task, we directly employ the expert data for InvertedPendulum-v2 in \cite{cheng2021guaranteed}. 

\begin{table*}[!tb]
\centering
    \caption{Hyperparameters for QIL algorithms.}
\label{table:hyperparameters}
\centering
\begin{tabular}{c c c c c c}
  \hline
  Environment    & Layers & Learning rates & Observables  & $\beta$ & Reward function \\
  \hline
  CartPole-v1 & 4 & $[0.07, 0.01, 0.07]$ & $[Z_0Z_1Z_2Z_3, -Z_0Z_1Z_2Z_3]$ & 1.0  & $-\log(D_\omega(s,a))$\\
  Acrobot-v1 & 5 & $[0.1, 0.01, 0.1]$ & $[Z_0, Z_0Z_1, Z_1]$ & 1.0 & $\log(1-D_\omega(s,a))$\\
  MountainCar-v0 & 6 & $[0.1, 0.01, 0.1]$ & $[Z_0, Z_0Z_1, Z_1]$ & 1.2  & $\log(1-D_\omega(s,a))$\\
  InvertedPendulum-v2 & 7 & $[0.07, 0.01, 0.001]$ & $[Z_0Z_1Z_2Z_3]$ & N/A  & $-\log(D_\omega(s,a))$\\
  \hline
\end{tabular}
\end{table*}

\begin{figure*}[!tb]
  \centering
      \subfigure[CartPole-v1]{
         \includegraphics[width=0.3\textwidth]{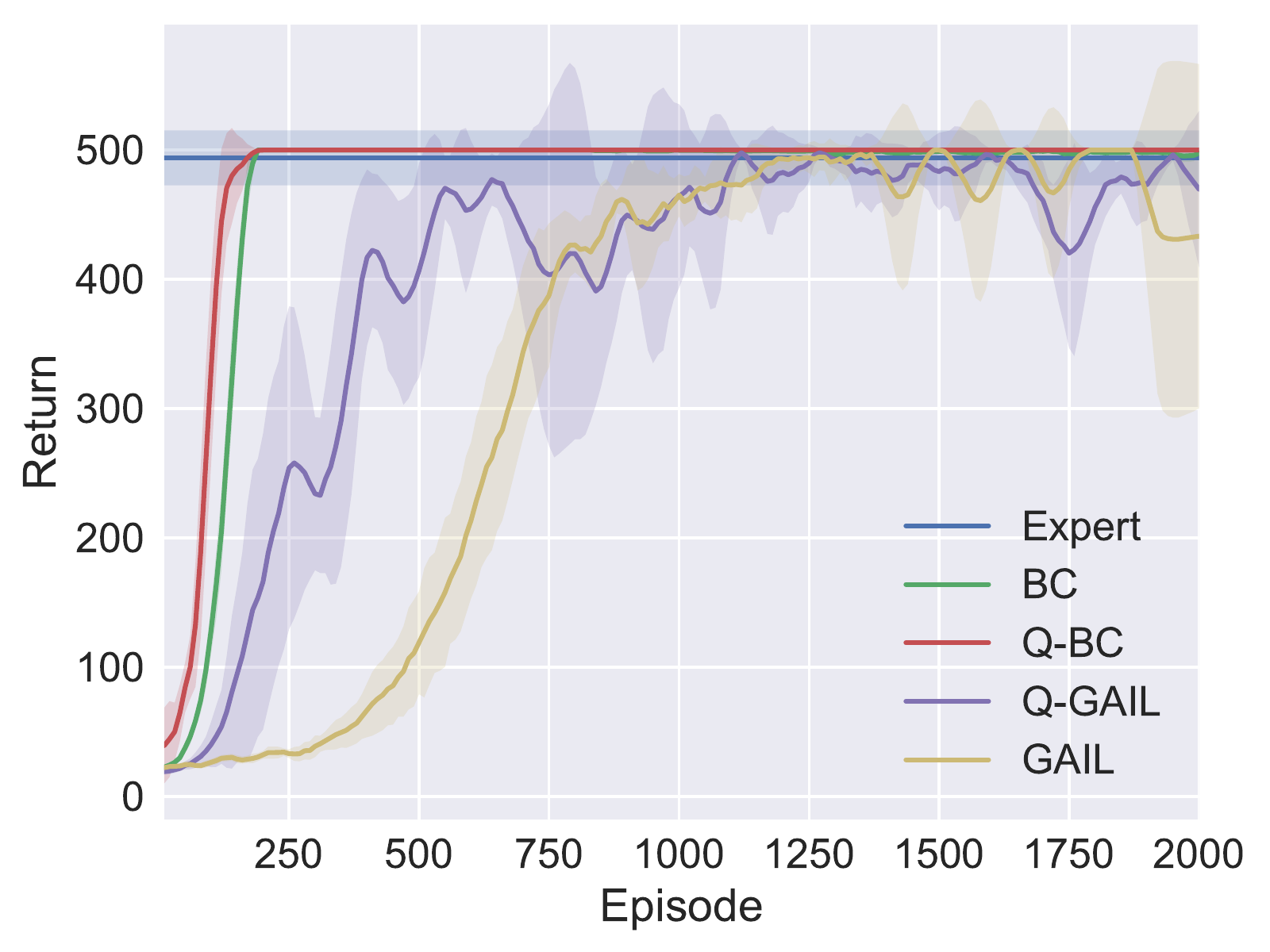}
      }
      \subfigure[Acrobot-v1]{
         \includegraphics[width=0.3\textwidth]{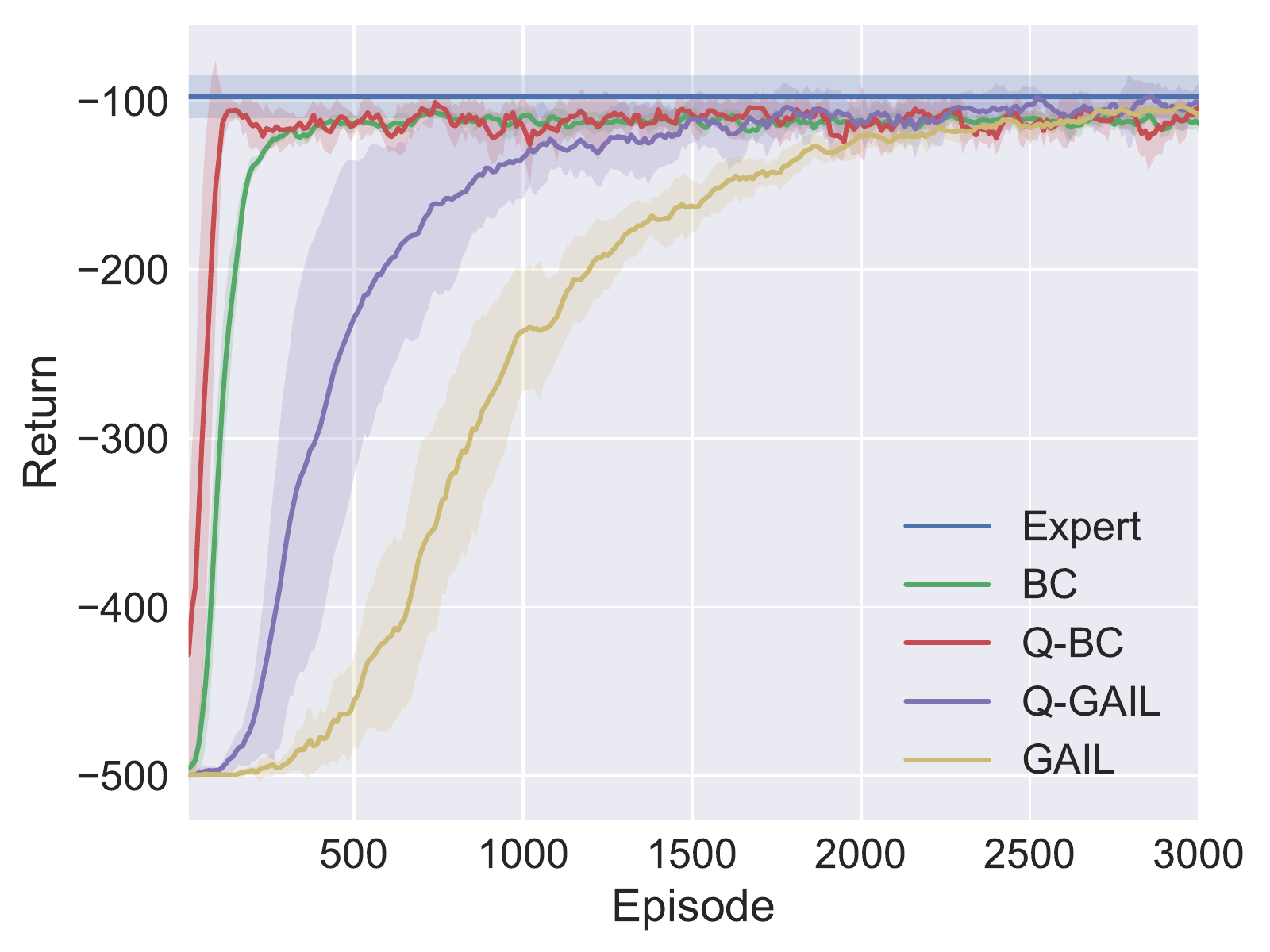}
      }
      \subfigure[MountainCar-v0]{
         \includegraphics[width=0.3\textwidth]{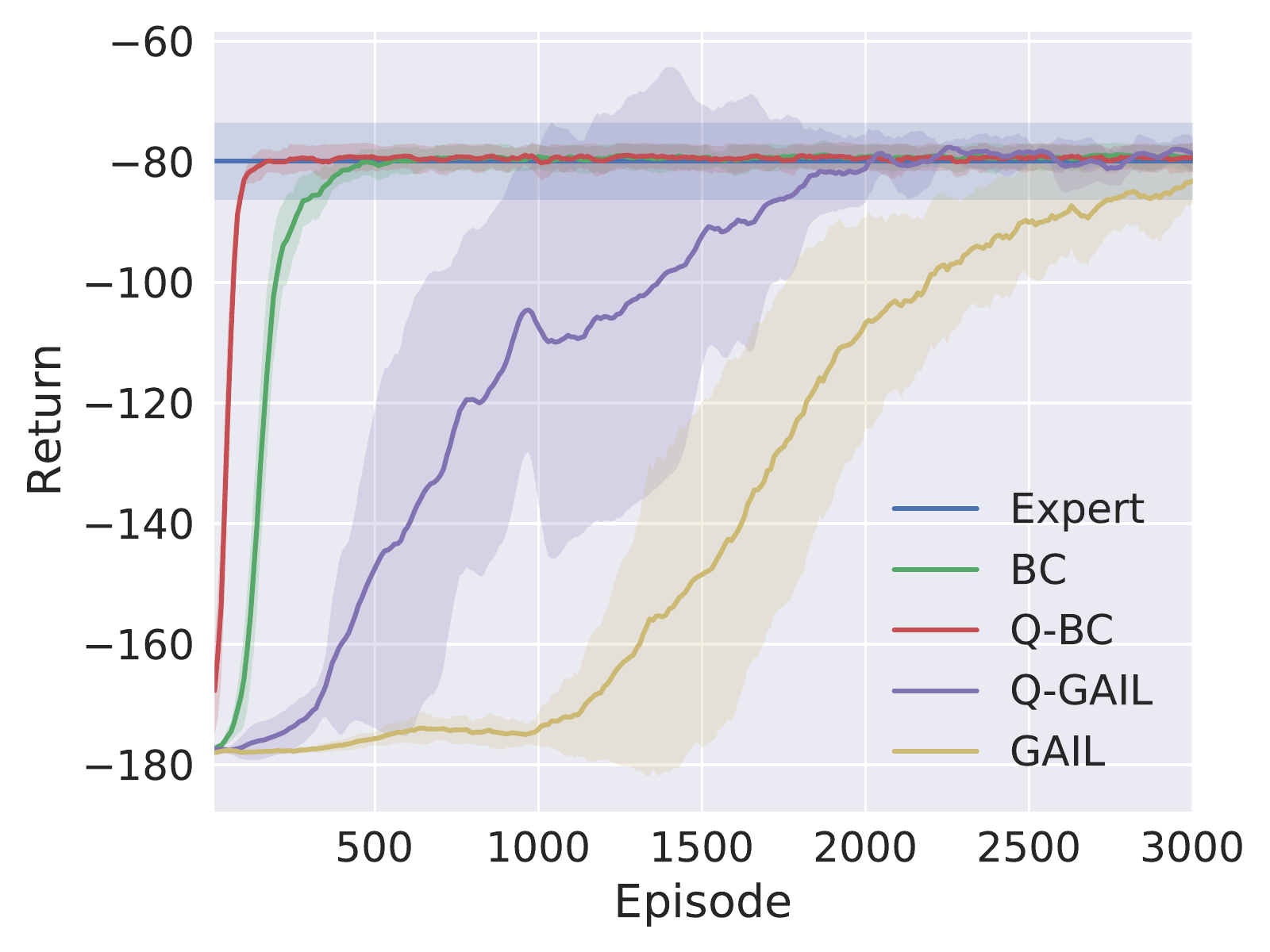}
      }
  \caption{Learning curves of Q-BC and Q-GAIL.}
  \label{fig:learning curves}
\end{figure*}

\subsection{Implementations and Hyperparameters}
We conduct experiments with Tensorflow Quantum~\cite{broughton2020tensorflow} and Cirq \cite{hancock2019cirq} to simulate quantum systems. The proposed QIL algorithms are implemented based on \cite{jerbi2021parametrized,dhariwal2017baselines,duan2016benchmarking}. Specifically, we follow \cite{jerbi2021parametrized} to construct VQCs and the training framework; the implementation of the discriminator is inspired by \cite{dhariwal2017baselines}; a linear \textit{baseline} is adopted from RLlab to reduce variance in IL \cite{duan2016benchmarking}.

We first introduce the shared hyperparamters of Q-BC and Q-GAIL for discrete-action tasks across different environments. The number of qubits for a VQC is the same as the state dimension of target environments as shown in Table \ref{table:gymspecfication}. The state bounds for CartPole-v1, Acrobot-v1, and MountainCar-v0 are $[2.4, 2.5, 0.21, 2.5]$, $[1, 1, 1, 1, 4\pi, 9\pi]$, and $[1.2, 0.07]$, respectively. For Q-BC, the batchsize is 2,000, which means that we sample 2,000 state-action pairs from expert data to train the policy at every iteration. Due to the compounding error, Q-BC could be more subtle to expert data. To better encode the input, we choose a smaller state bound according to the average and extreme value of expert data. The state bounds for CartPole-v1 and Acrobot-v1 in Q-BC are [1.0, 1.0, 1.0, 1.0], [0.5, 0.5, 0.5, 0.5, 3.0, 6.0], respectively. We also select local quantum observables, such as $[Z_0, Z_1]$ for CartPole-v1 and $[Z_0, Z_1, Z_2]$ for Acrobot-v1, to reduce the impact of barren plateaus. Furthermore, we use the commonly-used uniform initialization $[0,2\pi]$ to initialize parameters rather than $[0,\pi]$ in \cite{jerbi2021parametrized}.

For Q-GAIL, we use a discount $\gamma=1$ and a batchsize of 10. Note that the batchsize here stands for 10 agent trajectories rather than 10 state-action pairs. The discriminator is constructed with DNNs, which consists of 2 hidden layers with 64 nodes at each layer and its learning rate is $3e^{-4}$. The remaining hyperparamters are listed in Table~\ref{table:hyperparameters}. The policy $\pi_\theta$ contains three parts of parameters: input scaling parameters $\lambda$, variational parameters $\phi$, and output scaling parameters $\nu$. We select different learning rates for different parts such that $\xi_\theta=[\xi_\lambda, \xi_\phi, \xi_\nu]$, which is presented in Table~\ref{table:hyperparameters}. Pauli Z gates are used to convert quantum states into classical states, and $Z_i$ represents the readout of a Pauli Z gate on the $i$th qubit. 

As for the counterparts, traditional BC and GAIL, their policies are based on DNNs and share a same structure as the discriminator \cite{dhariwal2017baselines}, in which the learning rates of the discriminator and policy both are $1e^{-3}$. 

When we employ Gaussian-VQC for continuous-action tasks, hyper-parameters used for training QIL are slight modified, which is shown in Table~\ref{table:hyperparameters}. Note that for Q-GAIL, we use a VQC with 5 re-uploading layers. The state bound is $[2 \ 2 \ 2 \ 2]$, while the additional learning rate for the newly introduced parameter $\sigma$ is set to 0.002. Besides, in Q-GAIL for continuous-action tasks, we replace the initial RL algorithm REINFORCE with proximal policy optimization (PPO) \cite{schulman2017proximal}. The value of PPO-related parameters are adopted from \cite{SpinningUp2018}, \emph{i.e.}, we set the clip ratio to 0.2 and the target KL divergence as 0.01. The remaining parameters are the same as those of QIL algorithms for discrete-action environments.

\subsection{Results}
\begin{figure}[!h]
  \centering
      \subfigure[Q-BC]{
         \includegraphics[width=0.20\textwidth]{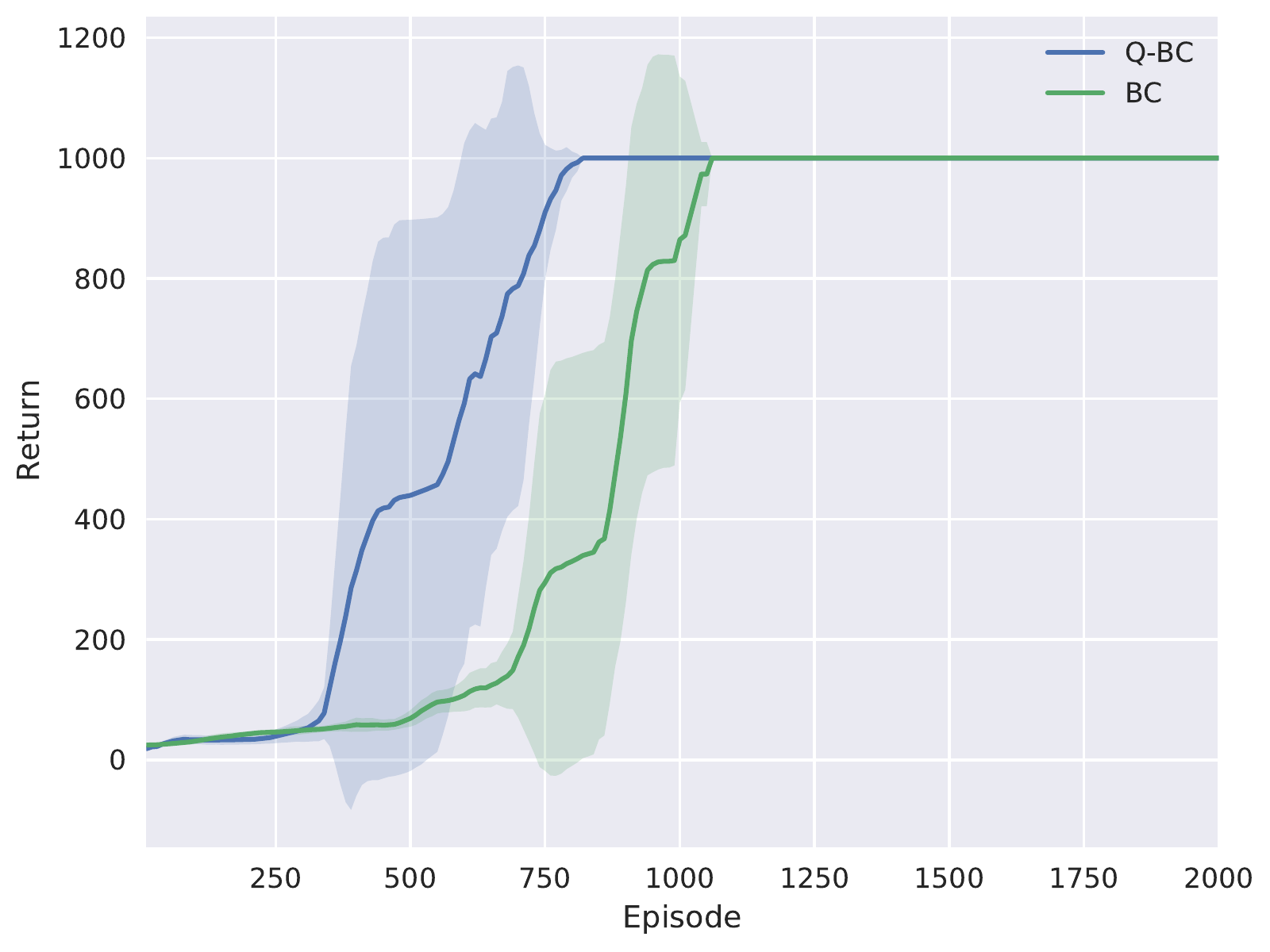}
      }
      \subfigure[Q-GAIL]{
         \includegraphics[width=0.2\textwidth]{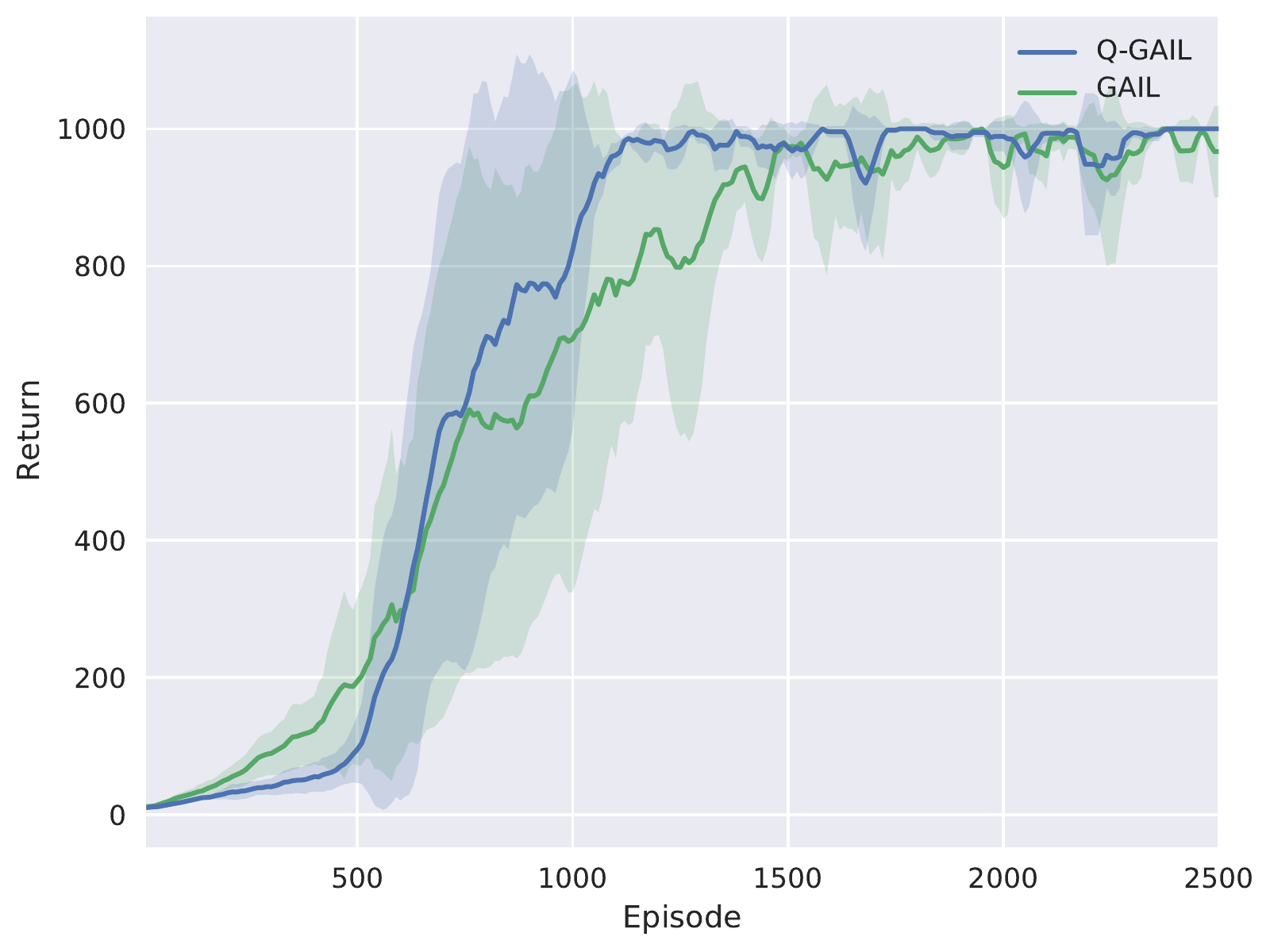}
      }
  \caption{Learning curves of Q-BC and Q-GAIL in InvertedPendulum-v2.}
  \label{fig:continu_learning_curves}
\end{figure}

In Fig. \ref{fig:learning curves}, we present the learning curves of our QIL algorithms and other counterparts for three control tasks. We also list the numerical results in Table \ref{table:numericalresults} for better contrast. Traditional BC and GAIL built with DNNs are selected as baselines. The performance is measured with episodic return. Each algorithm is run with 5 random seeds to obtain a relatively robust evaluation. Experimental results of Q-BC and Q-GAIL on the continuous task are presented in Fig. \ref{fig:continu_learning_curves}. Note that, in IL, agents have no access to the true rewards, and true rewards only serve as a performance metric \cite{ho2016generative}. Similar to \cite{jerbi2021parametrized, skolik2021quantum}, we employ the number of episodes to be the $x$-axis. A difference is that the $x$-axis for BC algorithms is test episodes, whereas for GAIL methods the $x$-axis represents the episodes that agents interact with the environment. Actually, BC-like IL algorithms do not require interactions with the environment during training. We choose to plot the performance of BC and Q-BC with respect to test episodes to better present the learning process. We can see from Figs. \ref{fig:learning curves} and \ref{fig:continu_learning_curves} that,
\begin{itemize}
    \item The proposed QIL algorithms, Q-BC and Q-GAIL, are able to successfully mimic experts from expert data and achieve expert-level performance. 
    \item Compared to BC and GAIL based on DNNs, no performance degradations are observed in the corresponding quantum versions, \emph{i.e.}, Q-BC and Q-GAIL. Besides, during training, QIL algorithms can achieve higher return with fewer samples and parameters.
    \item Q-BC learns faster and performs more stable than Q-GAIL across different environments.
\end{itemize}

Although both QIL and traditional IL algorithms can satisfactorily imitate experts, QIL algorithms possess some unique merits compared against the classical ones implemented with DNNs, \emph{i.e.}, QIL algorithms have the potential to be run on quantum computers and thus enjoy the quantum speed-up, which is further discussed in Subsection \ref{subsec:quantumadvantage}.

\begin{table*}[!h]
\centering
    \caption{Numerical performance of compared algorithms. We compare the performance of Q-BC and BC, as well as that of Q-GAIL and GAIL respectively. We use boldface type to indicate the better-performing method.}
\label{table:numericalresults}
\centering
\begin{tabular}{l c c c c c}
  \hline
  Environment   & BC & \textbf{Q-BC(Ours)} & GAIL  & \textbf{Q-GAIL(Ours)} & \\
  \hline
  CartPole-v1 & 497.6$\pm$1.3 & \textbf{500.0$\pm$0.0} & 433.4$\pm$133.1 & \textbf{469.7$\pm$60.5} \\
  Acrobot-v1 & -113.4$\pm$5.6 & \textbf{-104.0$\pm$5.2} & -108.2$\pm$2.5 & \textbf{-101.3$\pm$12.3} \\
  MountainCar-v0 & \textbf{-79.3$\pm$2.3} & \textbf{-79.3$\pm$2.5} & -83.1$\pm$3.6 & \textbf{-78.6$\pm$2.8} \\
  InvertedPendulum-v2 & \textbf{1000.0$\pm$0.0} & \textbf{1000.0$\pm$0.0} & 966.8$\pm$66.5 & \textbf{1000.0$\pm$0.0} \\
  \hline
\end{tabular}
\end{table*}

\begin{figure*}[!tb]
  \centering
      \subfigure[Q-BC]{
         \includegraphics[width=0.25\textwidth]{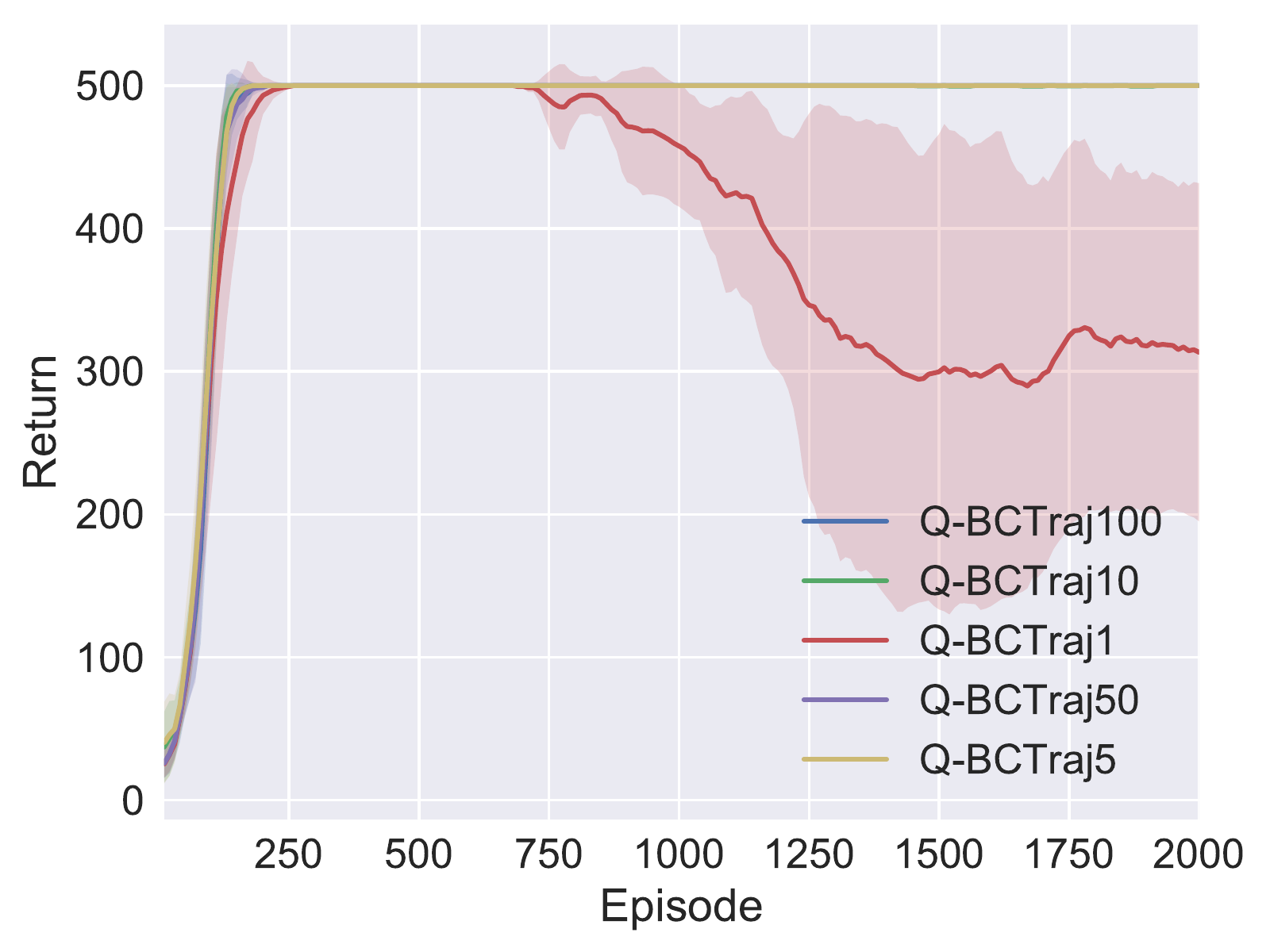}
         \includegraphics[width=0.25\textwidth]{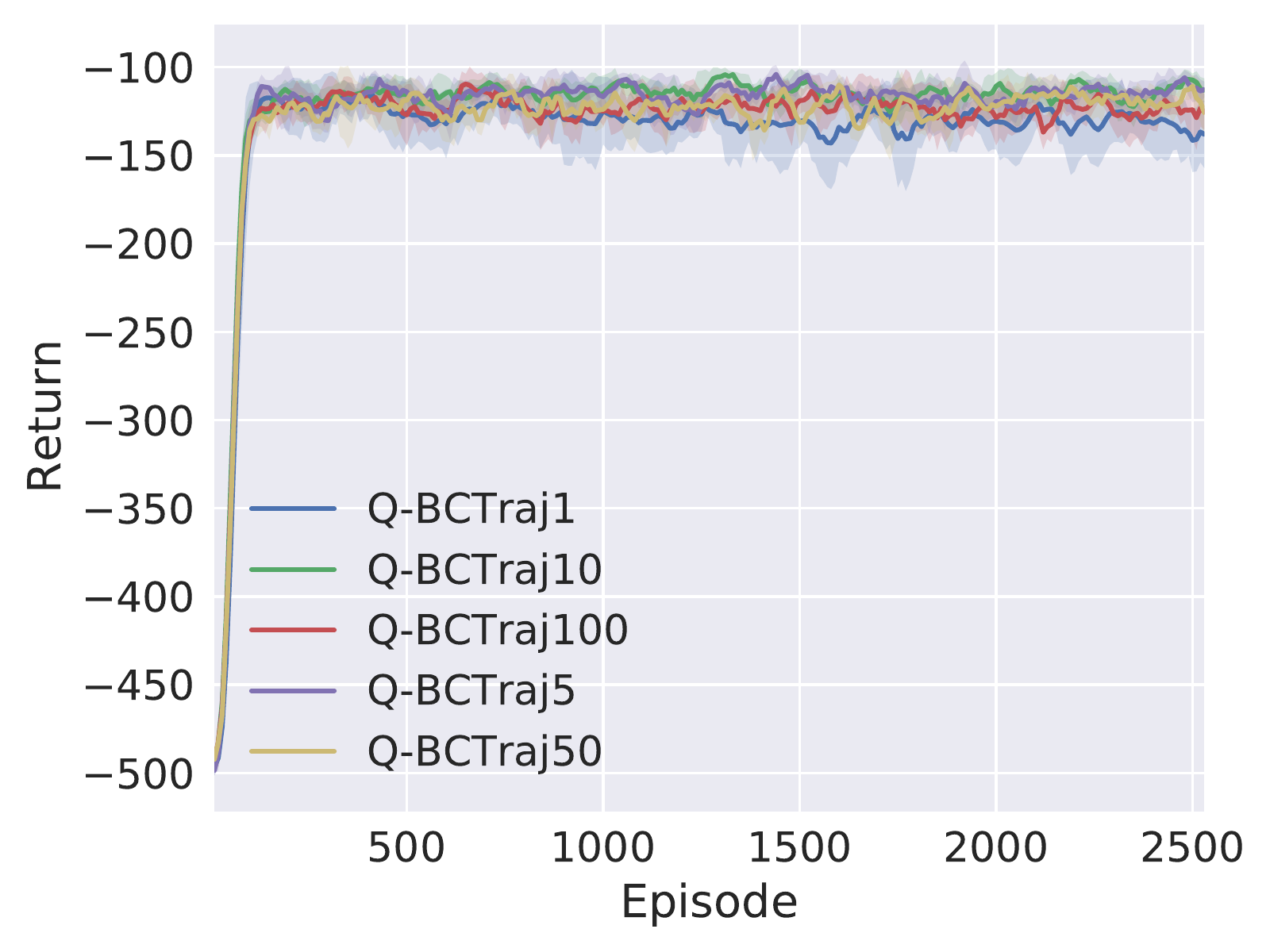}
         \includegraphics[width=0.25\textwidth]{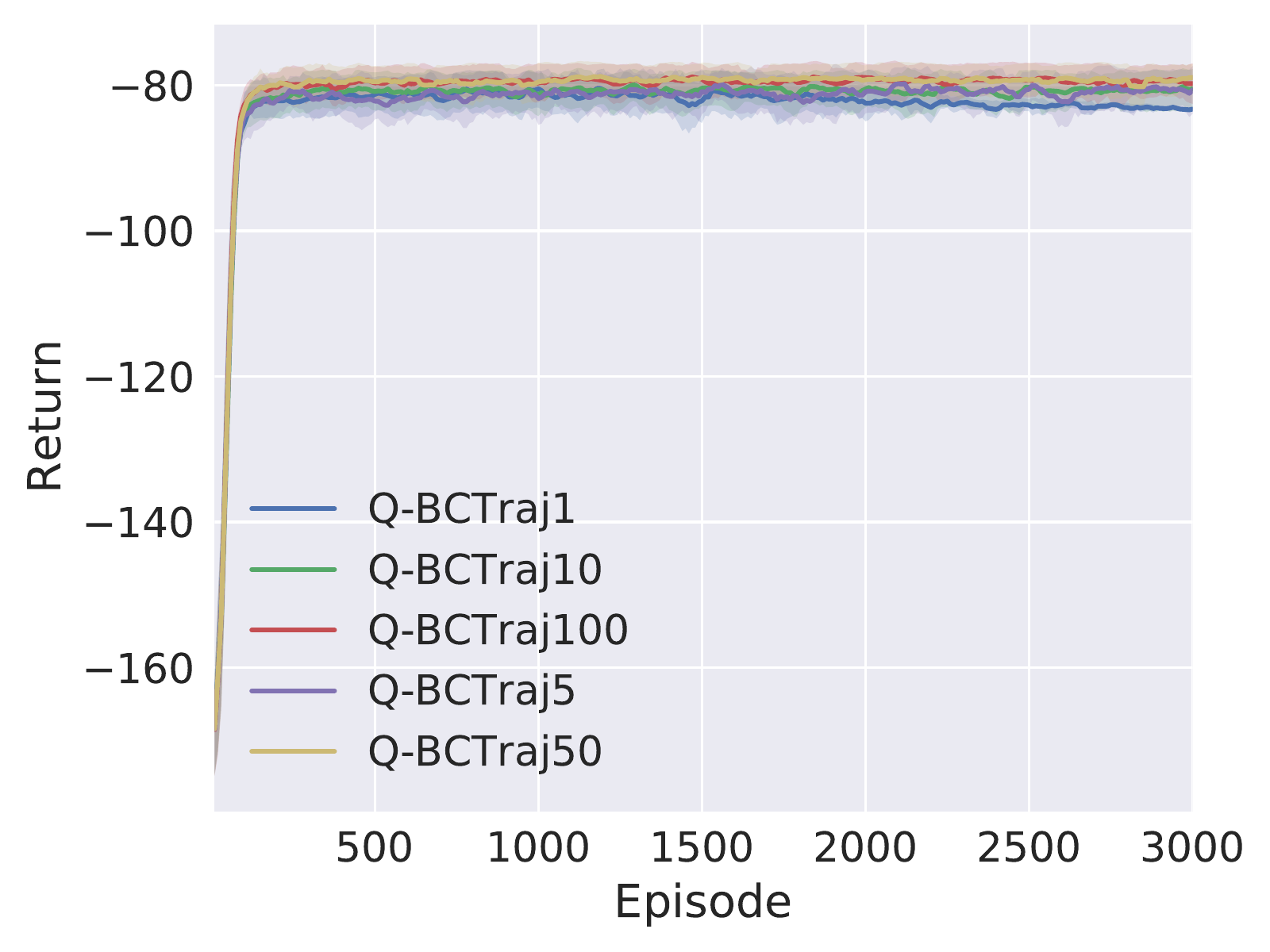}
      } 
      \subfigure[Q-GAIL]{
         \includegraphics[width=0.25\textwidth]{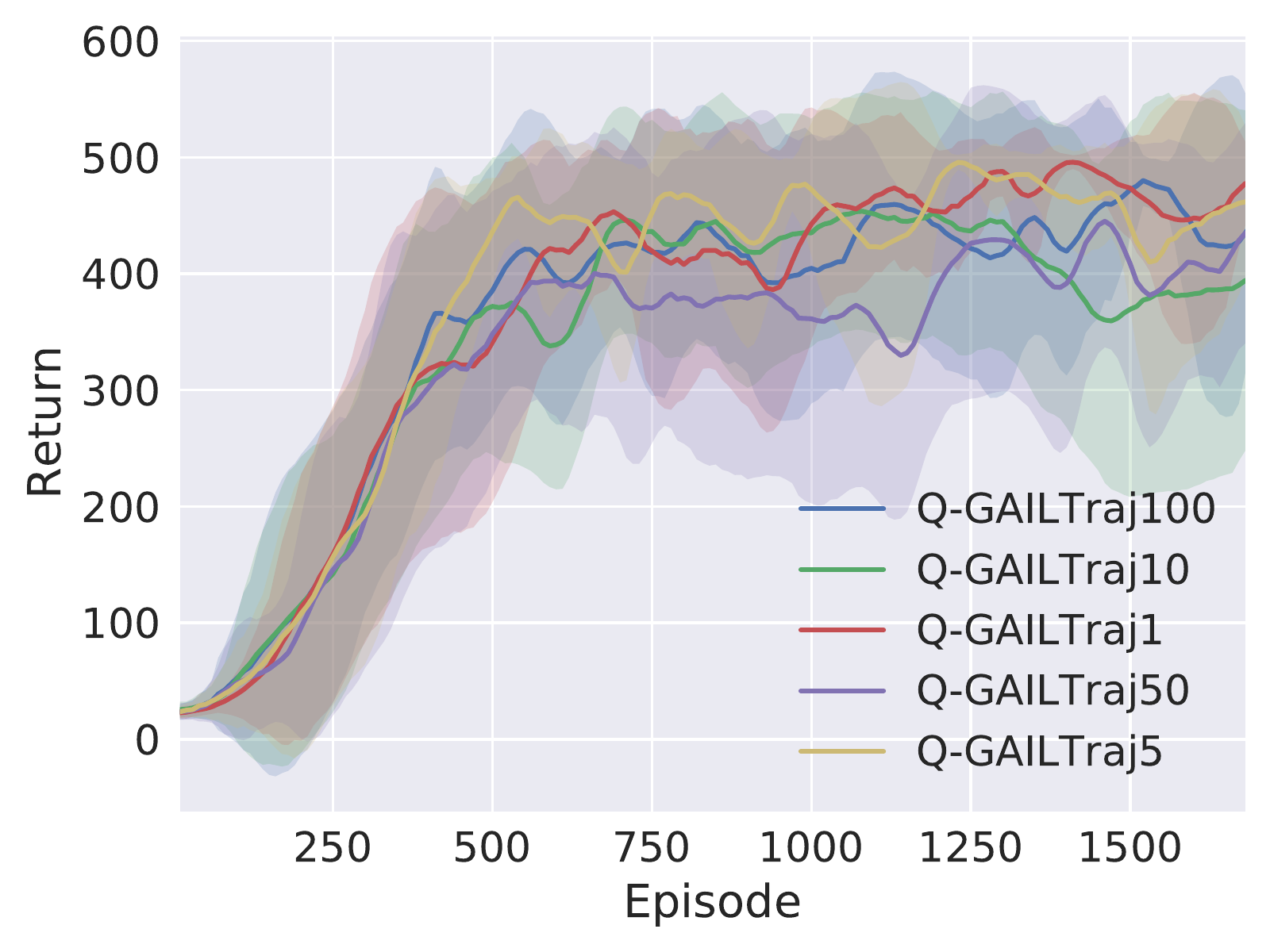}
         \includegraphics[width=0.25\textwidth]{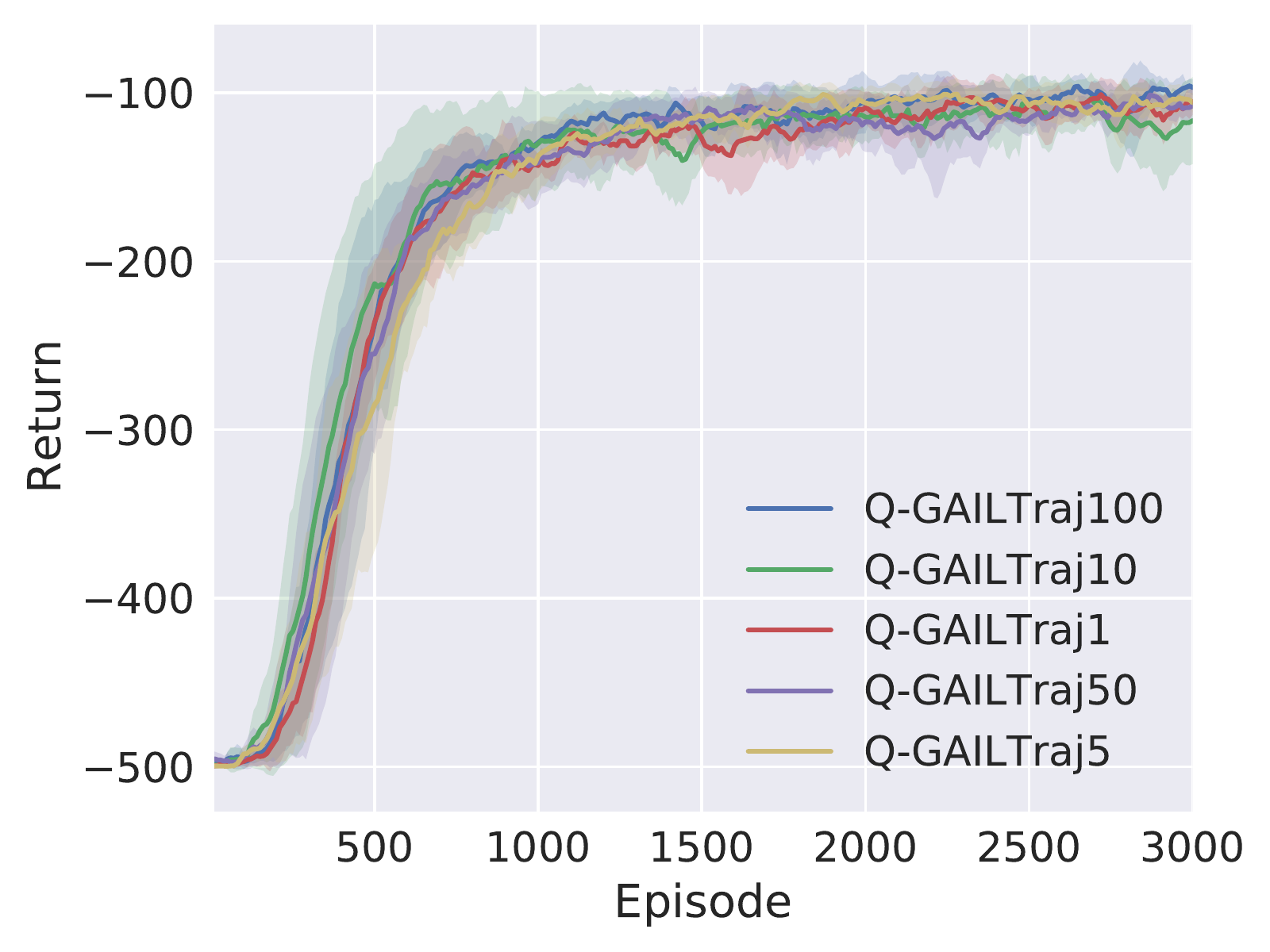}
         \includegraphics[width=0.25\textwidth]{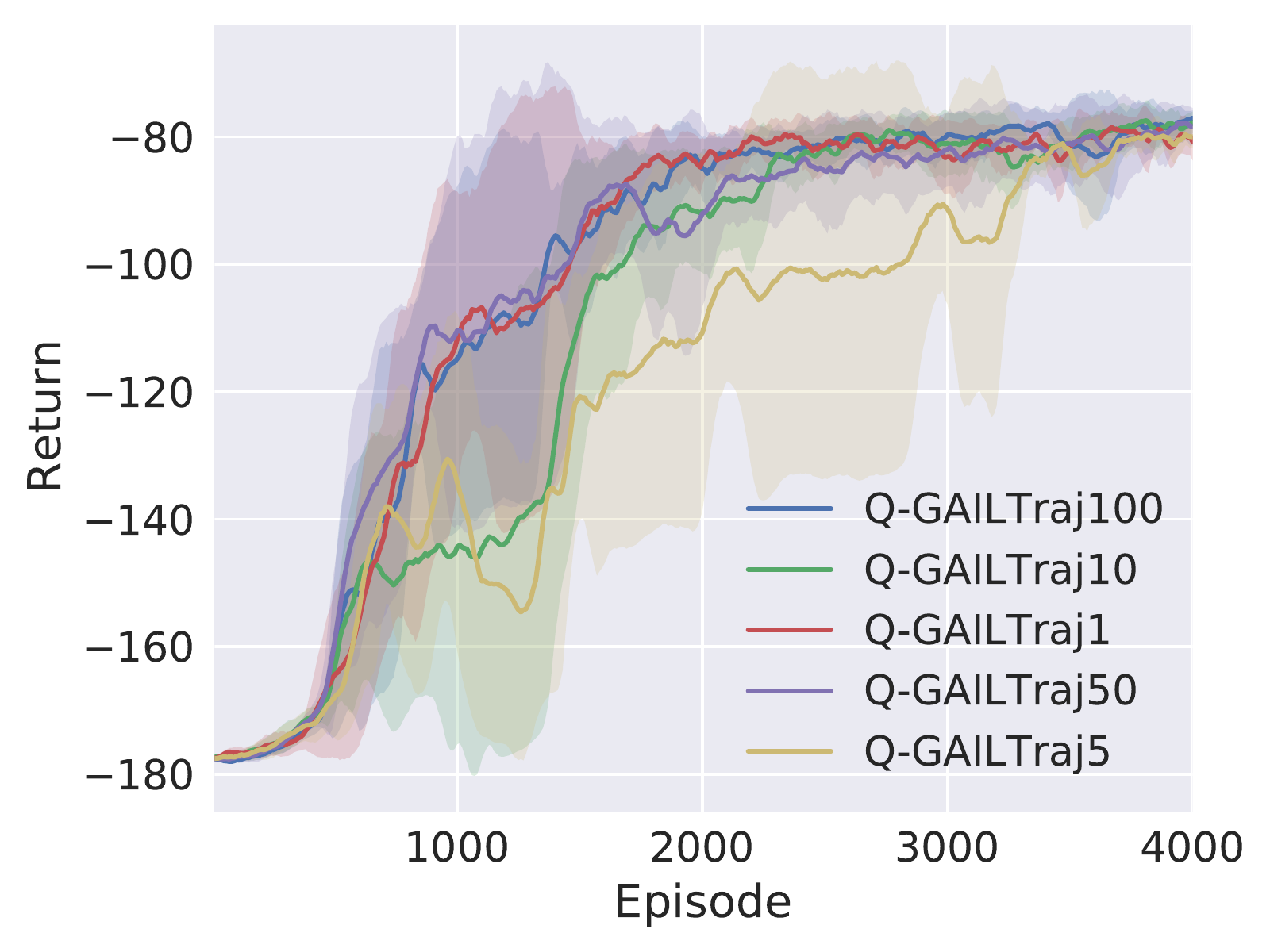}
      }
  \caption{Impact of the number of expert trajectories on the performance of Q-BC and Q-GAIL. The curves in the first, second, and third columns are obtained in CartPole-v1, Acrobot-v1, and MountainCar-v0, respectively.}
  \label{fig:numoftraj}
\end{figure*}

\subsection{Ablation Studies}\label{subsec:ablation}
In this subsection, we mainly ablate the common factors that could affect the performance of developed QIL algorithms, which can provide further insights and guide the employment of QIL. In general, three crucial factors are investigated, \emph{i.e.}, number of expert trajectories, VQC structures, and spectral normalization. Besides, we also test the performance of Q-GAIL with quantum discriminators.

\subsubsection{Number of Expert Trajectories}
In this part, we present experiments to investigate the impact of the number of expert data used for QIL, which are presented in Fig. \ref{fig:numoftraj}. The QIL algorithms, Q-BC and Q-GAIL, are trained in CartPole-v1, Acrobot-v1, and MountainCar-v0 with different number of expert trajectories (1, 5, 10, 50, 100). For CartPole-v1, corresponding to those expert trajectories, the numbers of state-action pairs are (499, 2499, 4999, 24663, 49329). The state-action pairs for Acrobot-v1 are (81, 479, 922, 4946, 9803), respectively, whereas the numbers for MountainCar-v0 are (119, 590, 1193, 5882, 11833), respectively. For experiments of Q-GAIL: in CartPole-v1, the observable used for quantum measurement is $[Z_0, Z_1]$; in MountainCar-v0, we choose a smaller $\beta=0.5$ to stimulate exploration to find better policies. The other hyperparameters in Q-BC are the same as those of Q-GAIL.

From Fig.~\ref{fig:numoftraj}, we can see that for Q-BC, more expert data help improve performance. In CartPole-v1, Q-BC can only achieve about half of the expert performance with one expert trajectory. On the contrary, Q-BC can perform as well as experts with 5 or more expert trajectories in CartPole-v1. Besides, the performance degradation of Q-BC in CartPole-v1 is more noticeable compared to that in the other two environments. The reason for this phenomenon is that the episode length is much longer in CartPole-v1, leading to larger compounding errors and overfitting with fewer expert data. 

\begin{figure}[!h]
  \centering
      \subfigure[Q-BC]{
         \includegraphics[width=0.23\textwidth]{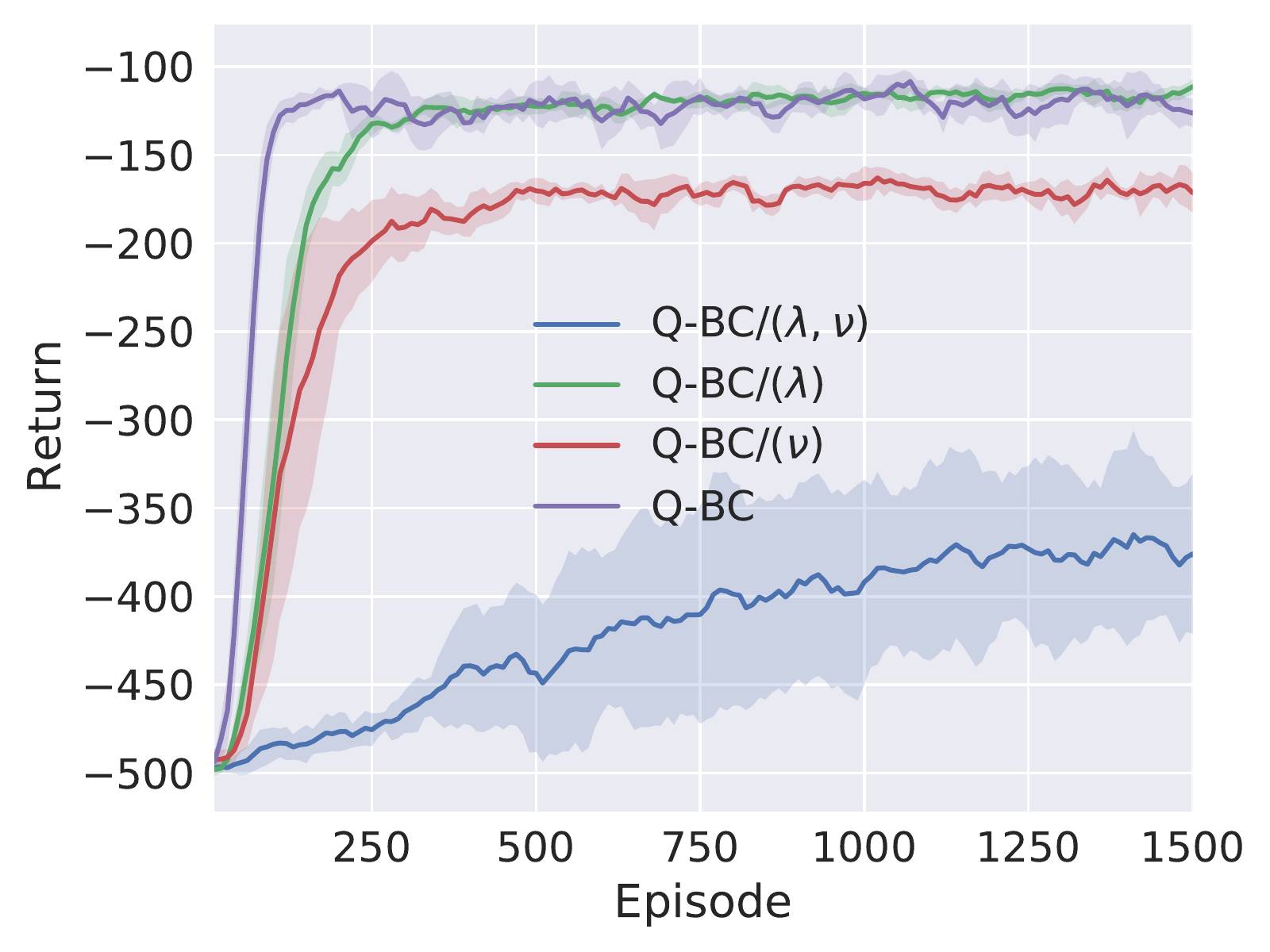}
         \includegraphics[width=0.23\textwidth]{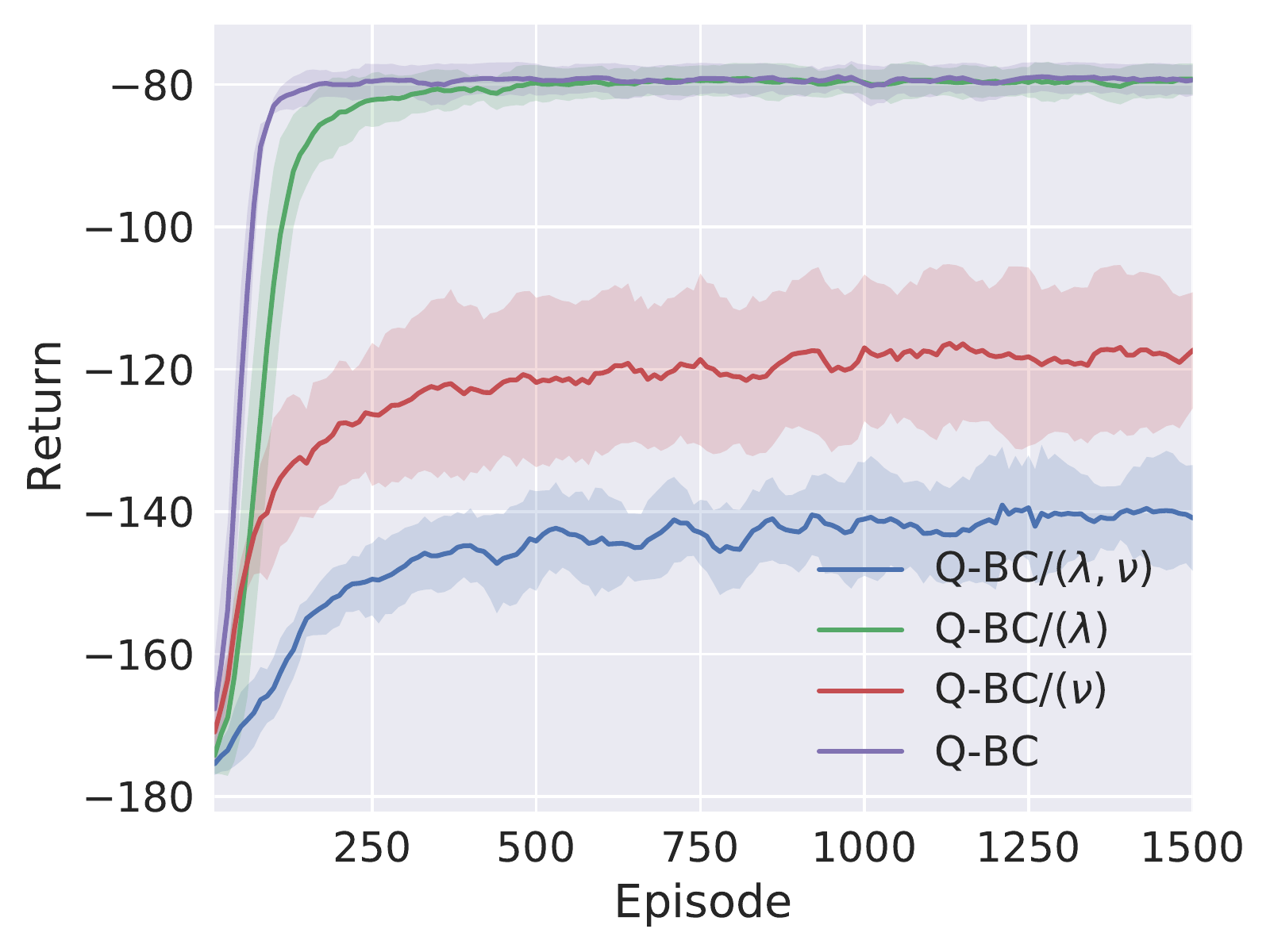}
      }
      \subfigure[Q-GAIL]{
         \includegraphics[width=0.23\textwidth]{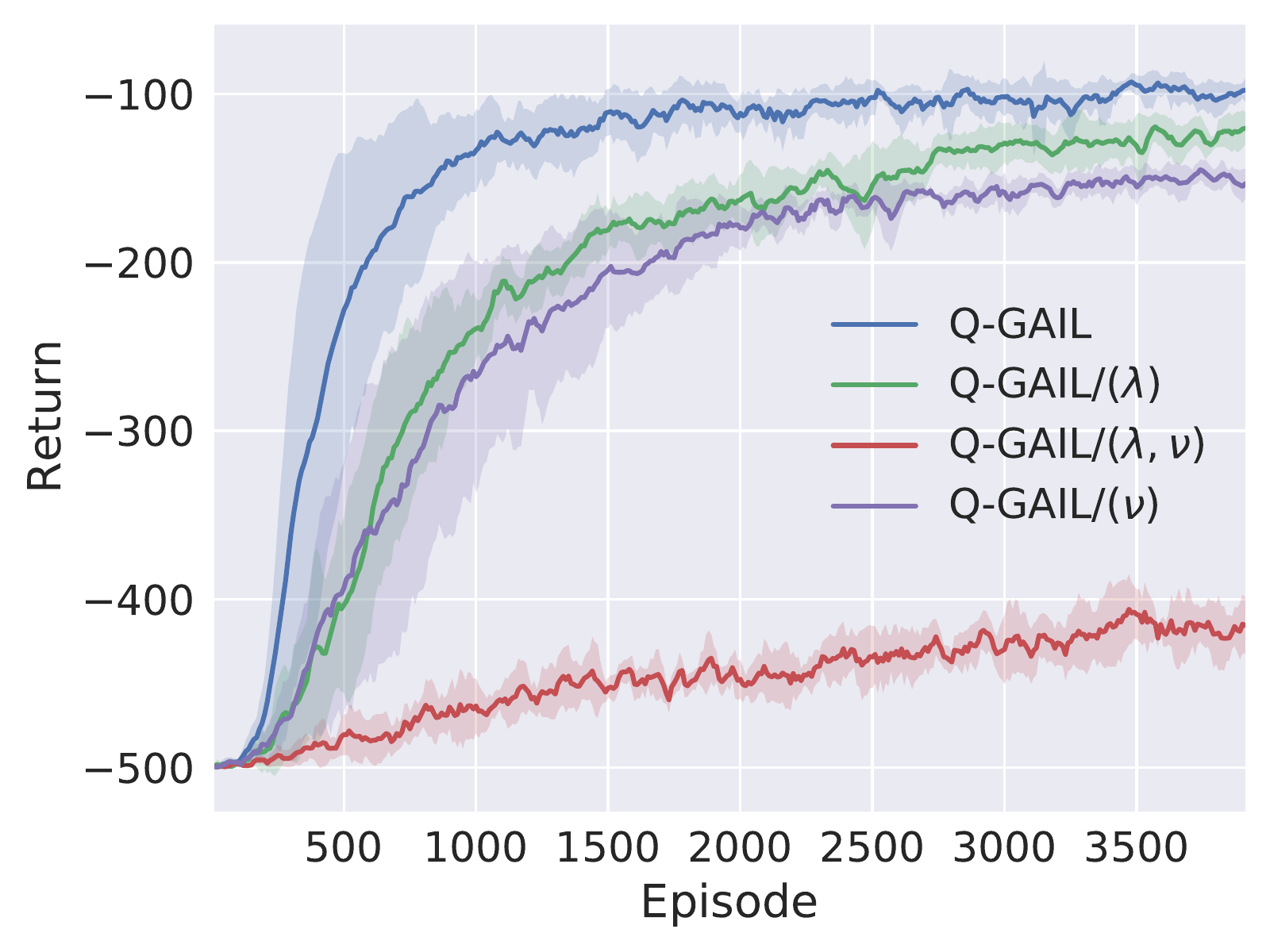}
         \includegraphics[width=0.23\textwidth]{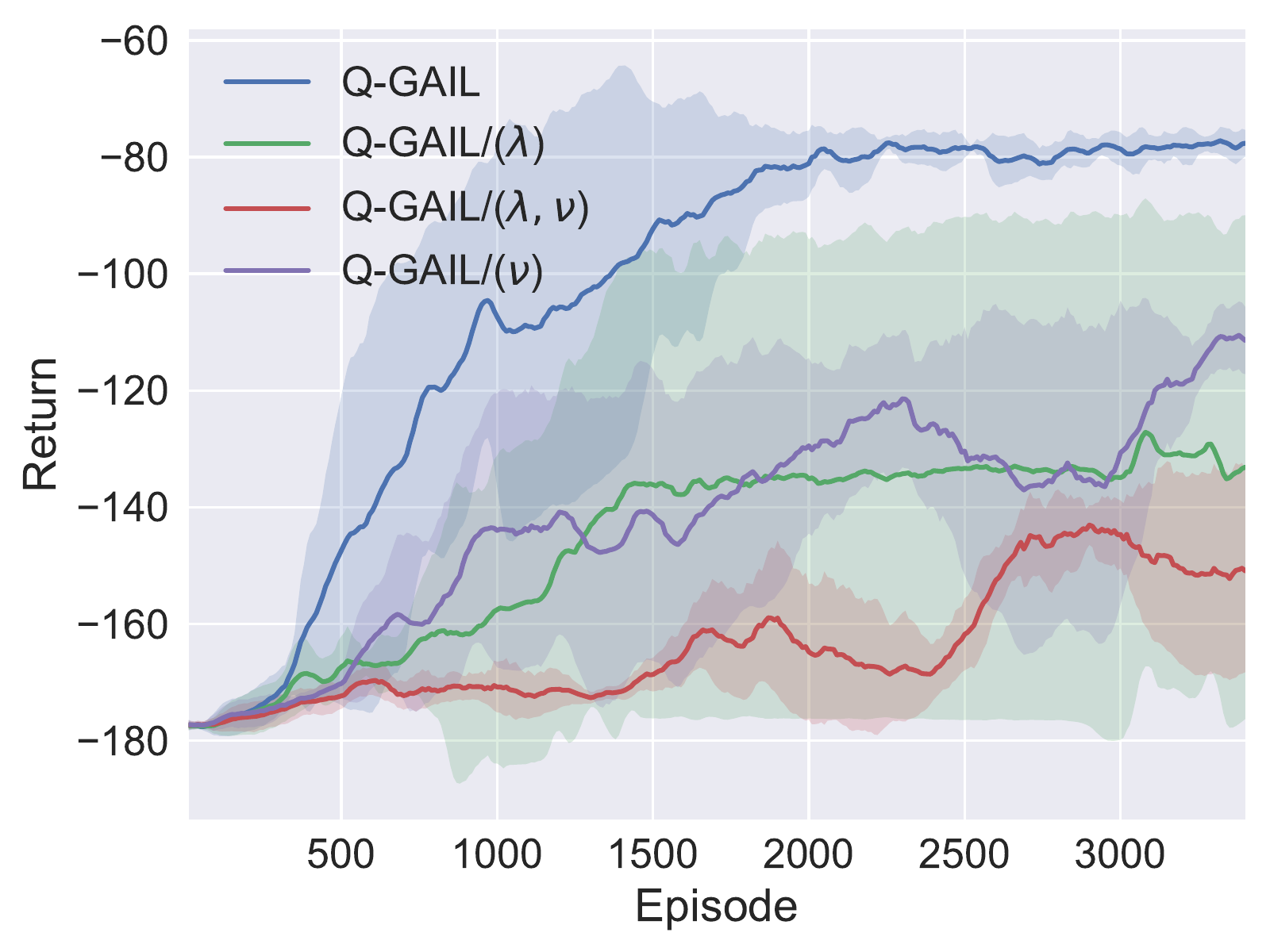}
      }
  \caption{Ablations of scaling parameters. $\lambda$ represents input scaling parameters, while $\nu$ stands for output scaling parameters. The legend Q-BC/$(x)$ or Q-GAIL/$(x)$ denote the corresponding version whose scaling parameter $x$ is disabled. }
  \label{fig:ablationscalingparams}
\end{figure}
\begin{figure}[!h]
  \centering
      \subfigure[Q-BC]{
         \includegraphics[width=0.23\textwidth]{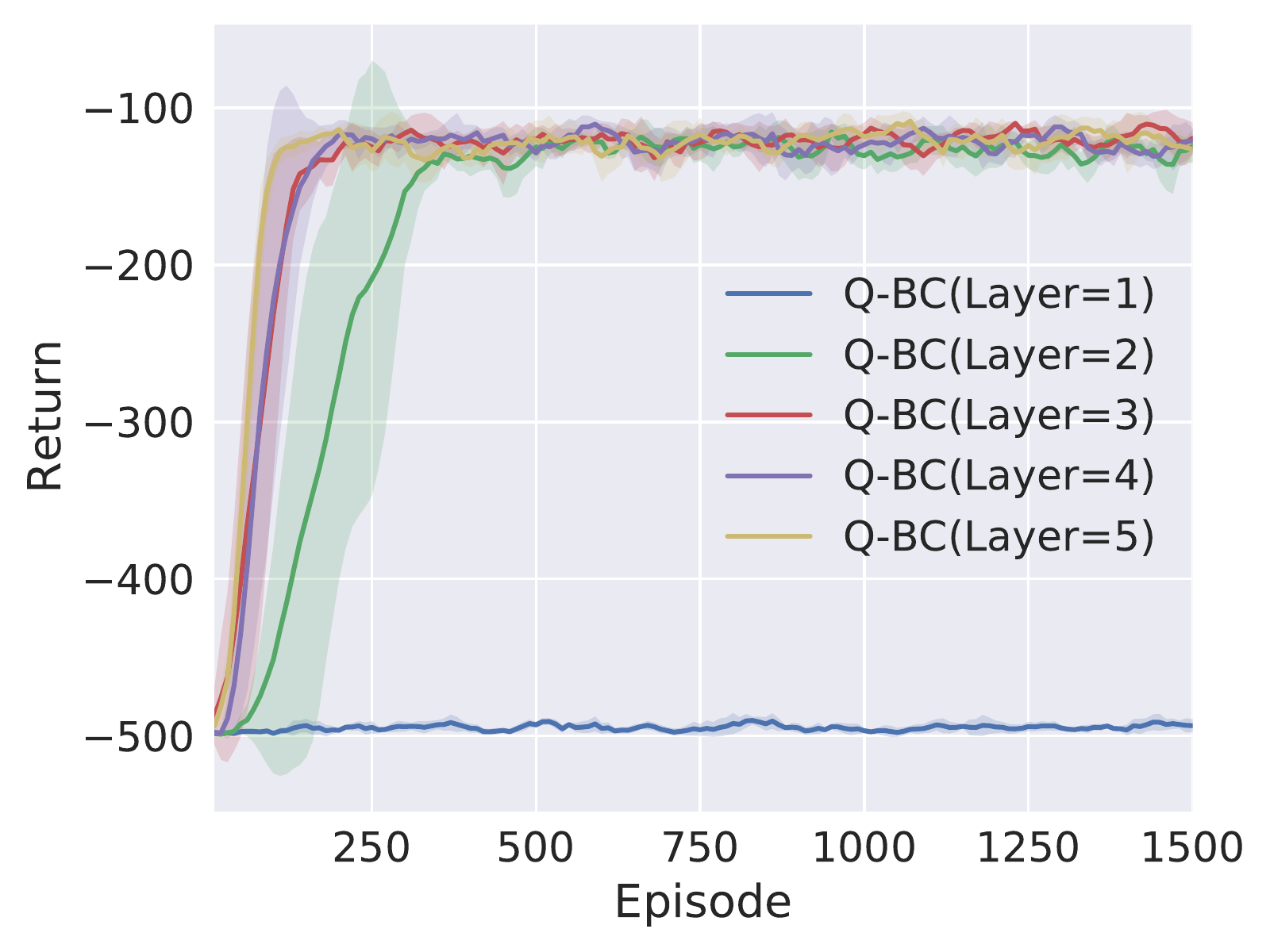}
         \includegraphics[width=0.23\textwidth]{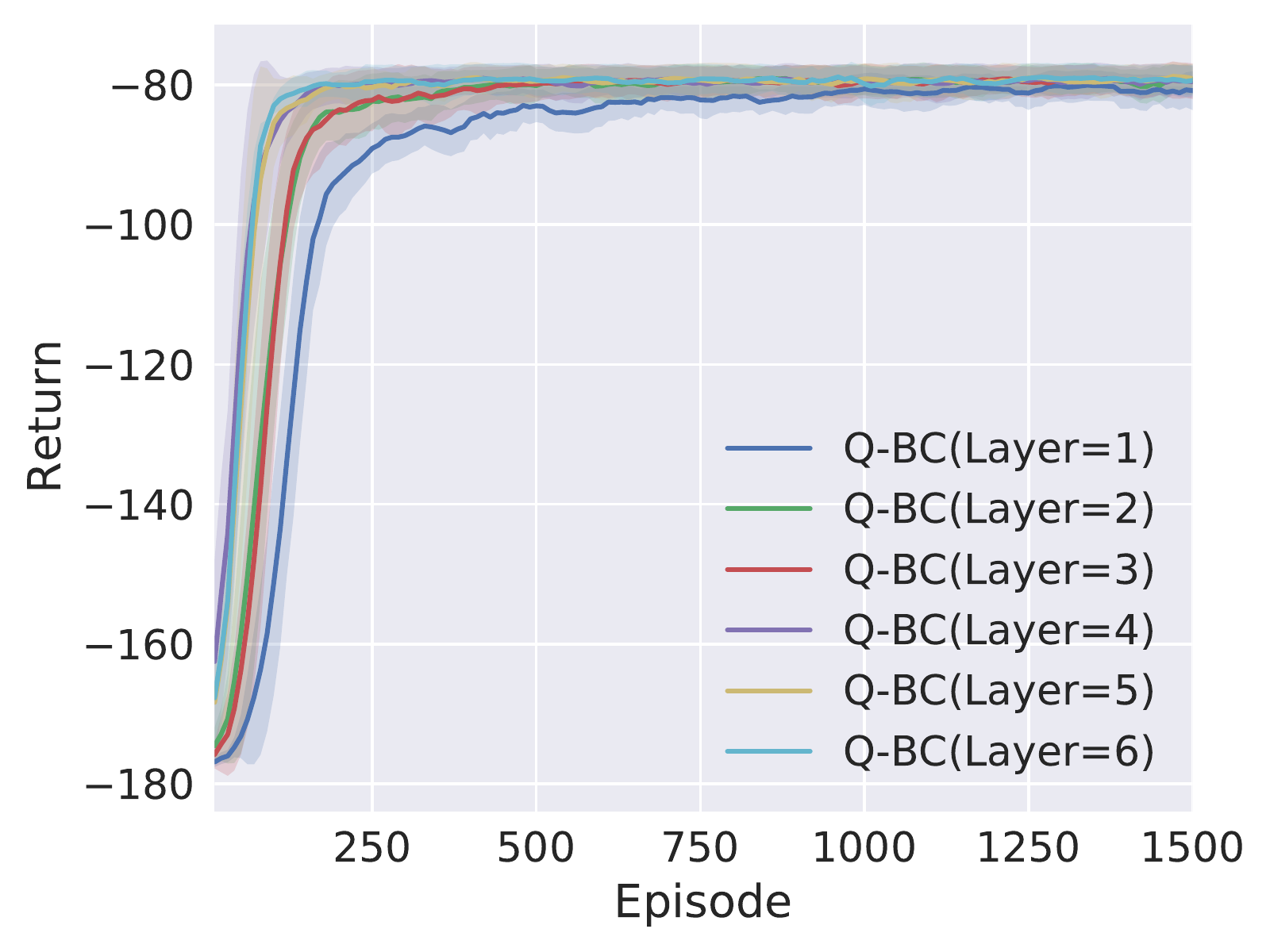}
      }
      \subfigure[Q-GAIL]{
         \includegraphics[width=0.23\textwidth]{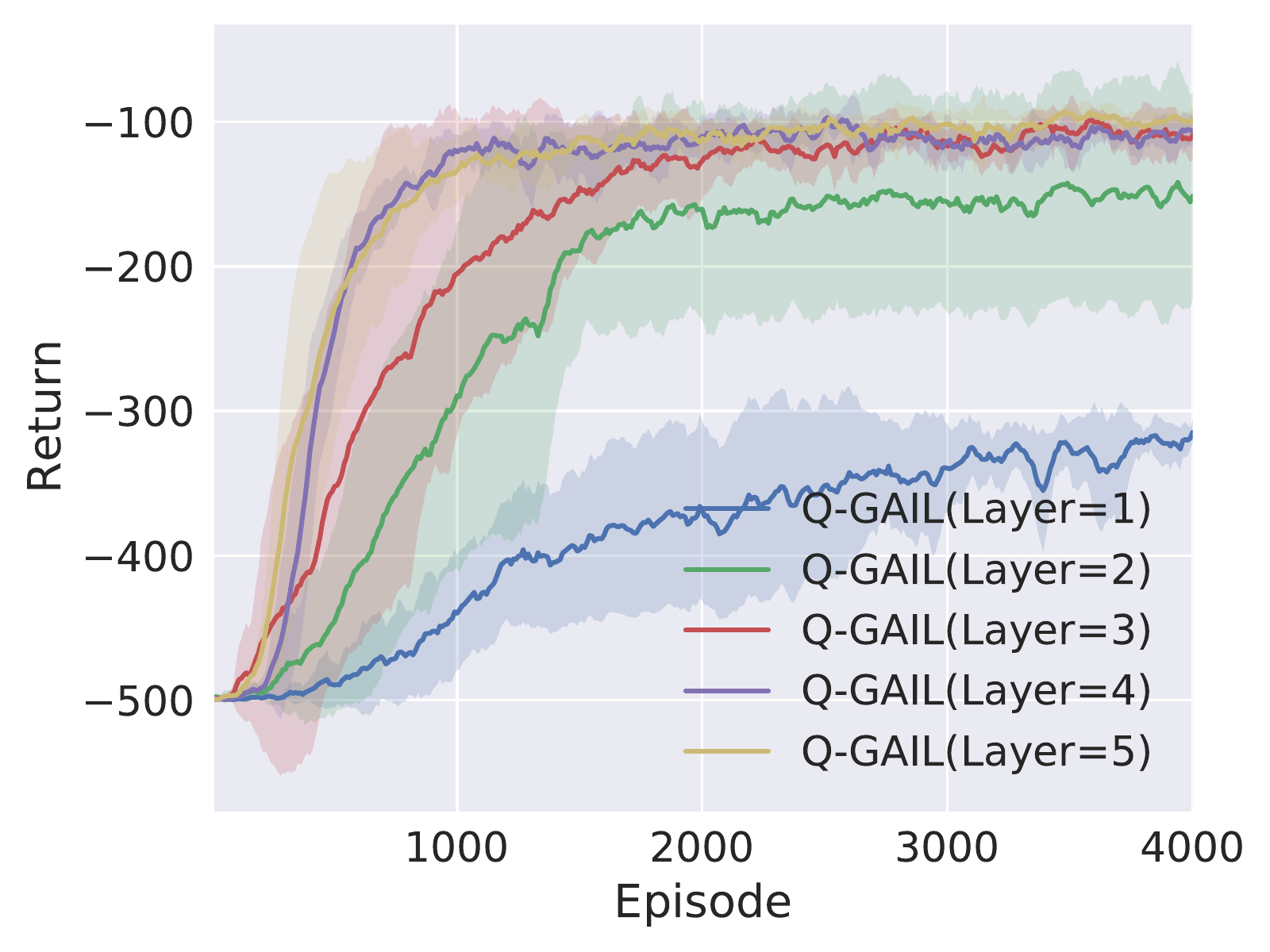}
         \includegraphics[width=0.23\textwidth]{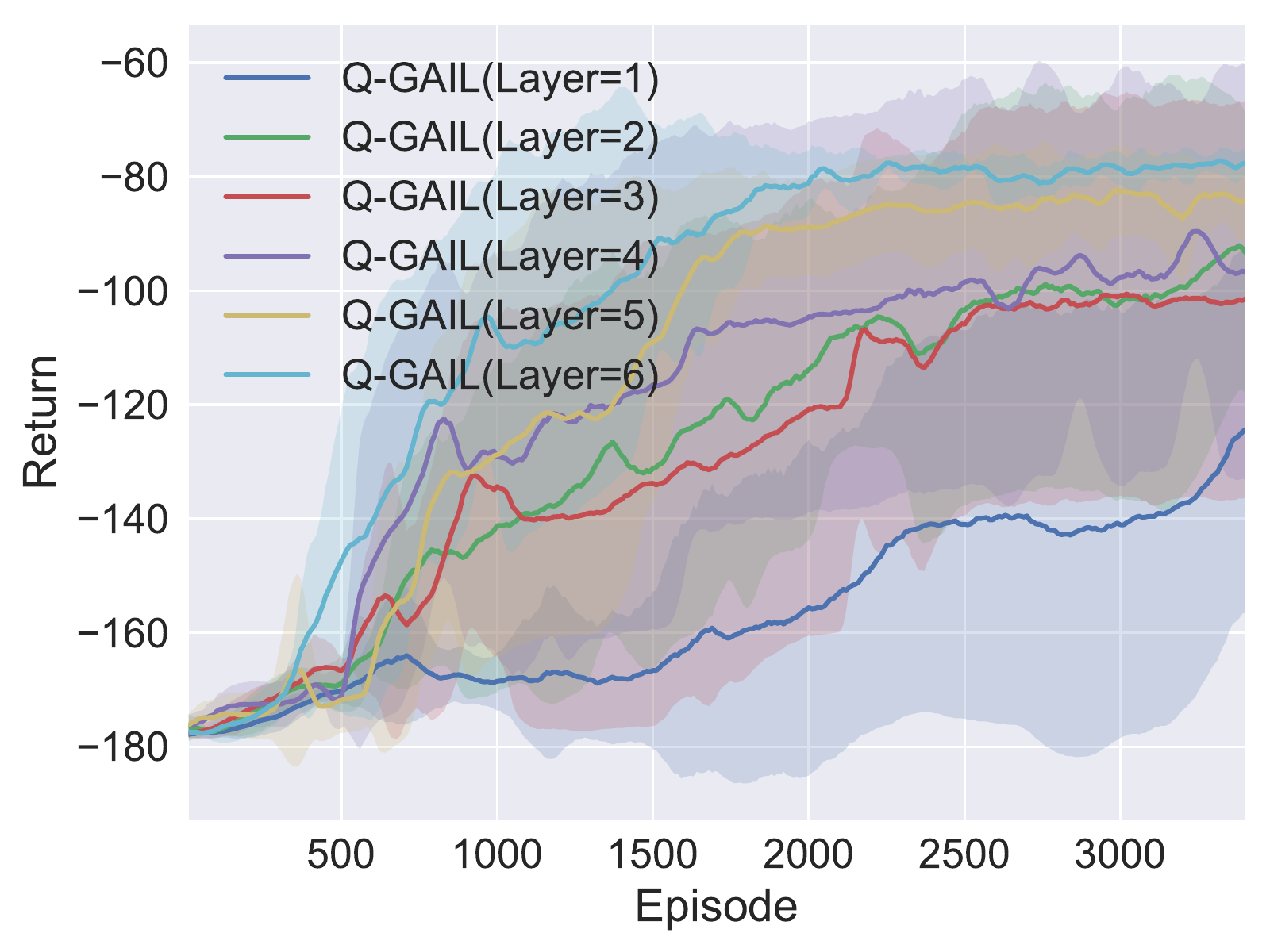}
      }
  \caption{Ablations of the number of layers of softmax-VQCs.}
  \label{fig:ablationlayers}
\end{figure}

In contrast, Q-GAIL is able to effectively replicate expert behaviors with only one expert trajectory, which means that Q-GAIL is significantly more data-efficient compared to Q-BC. We also observe that increasing the number of expert data does not help improve the performance of Q-GAIL. In addition, Q-GAIL exhibits small fluctuations in the performance of CartPole-v1. The unstable characteristic of CartPole-v1, the oscillating property of VQCs \cite{jerbi2021parametrized}, and the adversarial IL scheme together contribute to the observed fluctuations. In the other two environments, Q-GAIL demonstrates considerably stable performance.

\begin{figure}[!t]
  \centering
      \subfigure[Q-BC]{
         \includegraphics[width=0.23\textwidth]{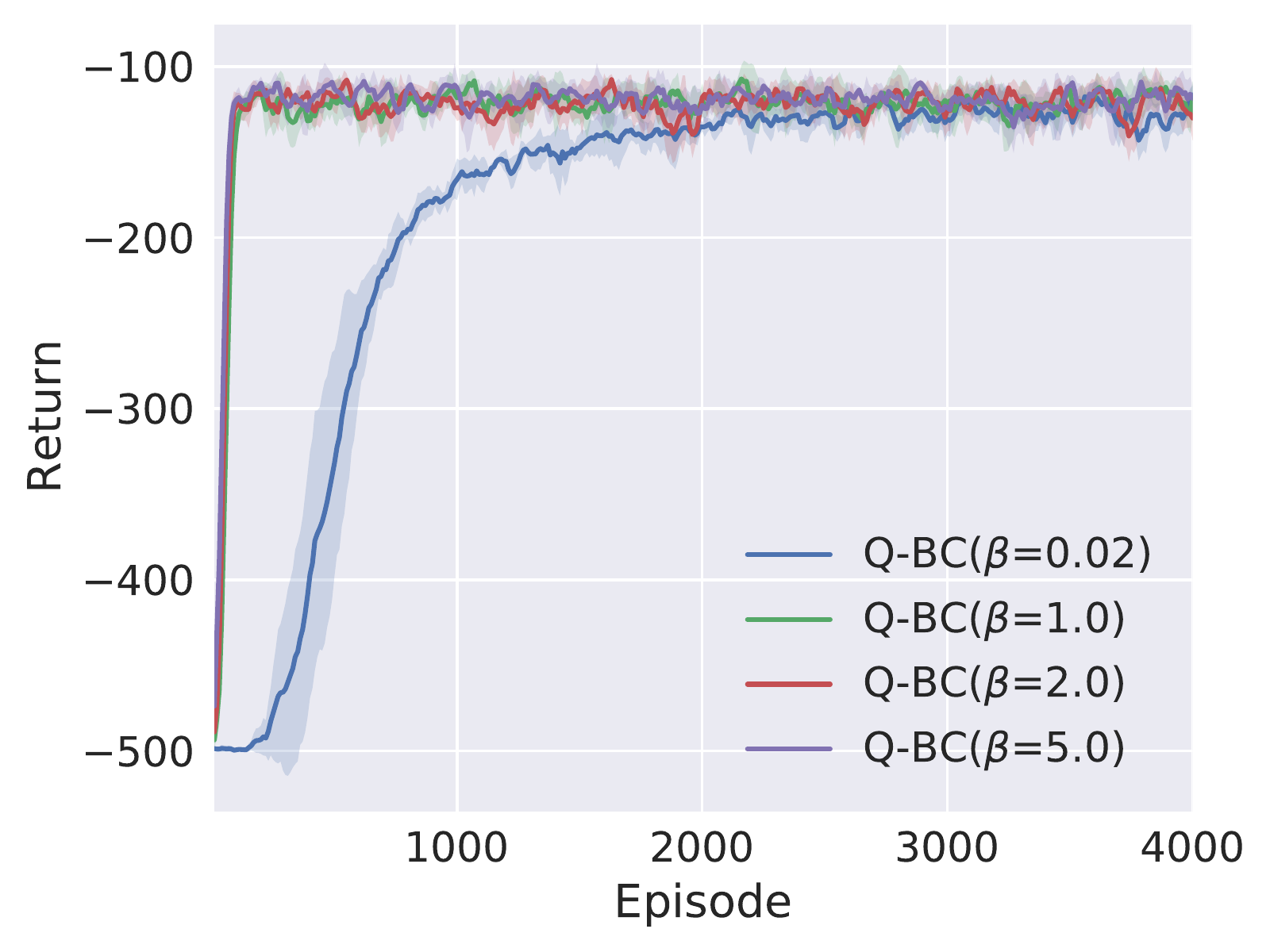}
         \includegraphics[width=0.23\textwidth]{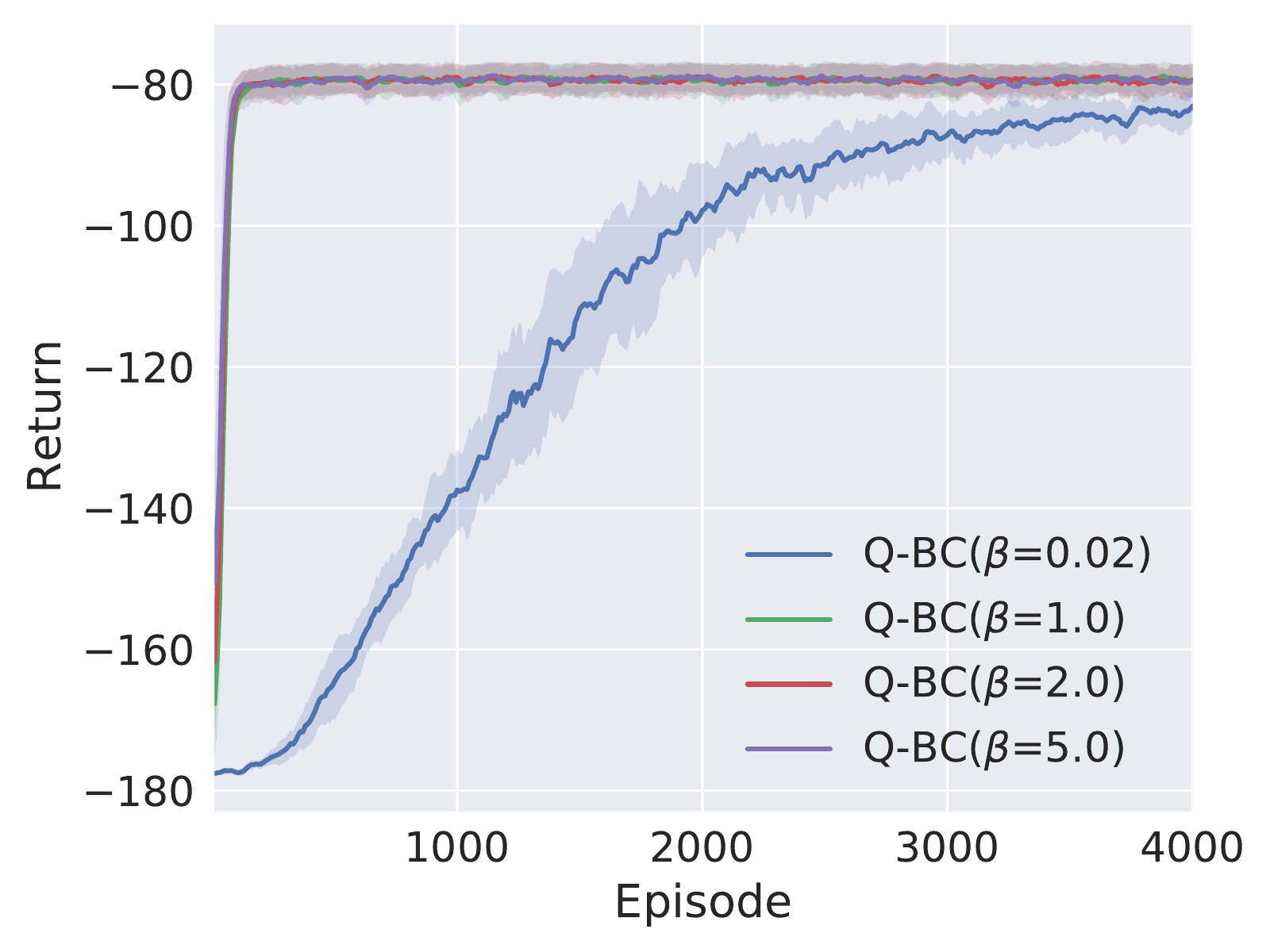}
      }
      \subfigure[Q-GAIL]{
         \includegraphics[width=0.23\textwidth]{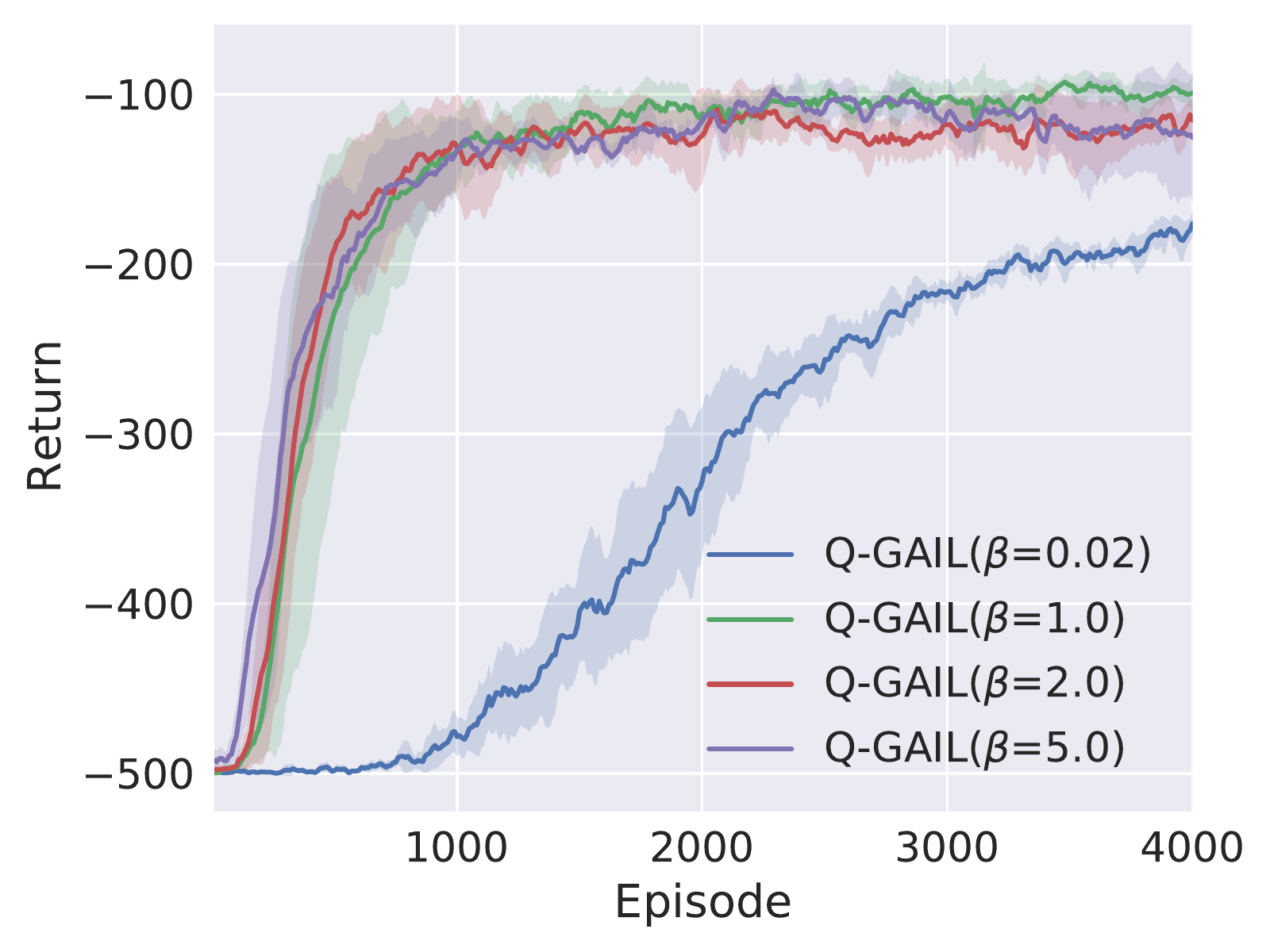}
         \includegraphics[width=0.23\textwidth]{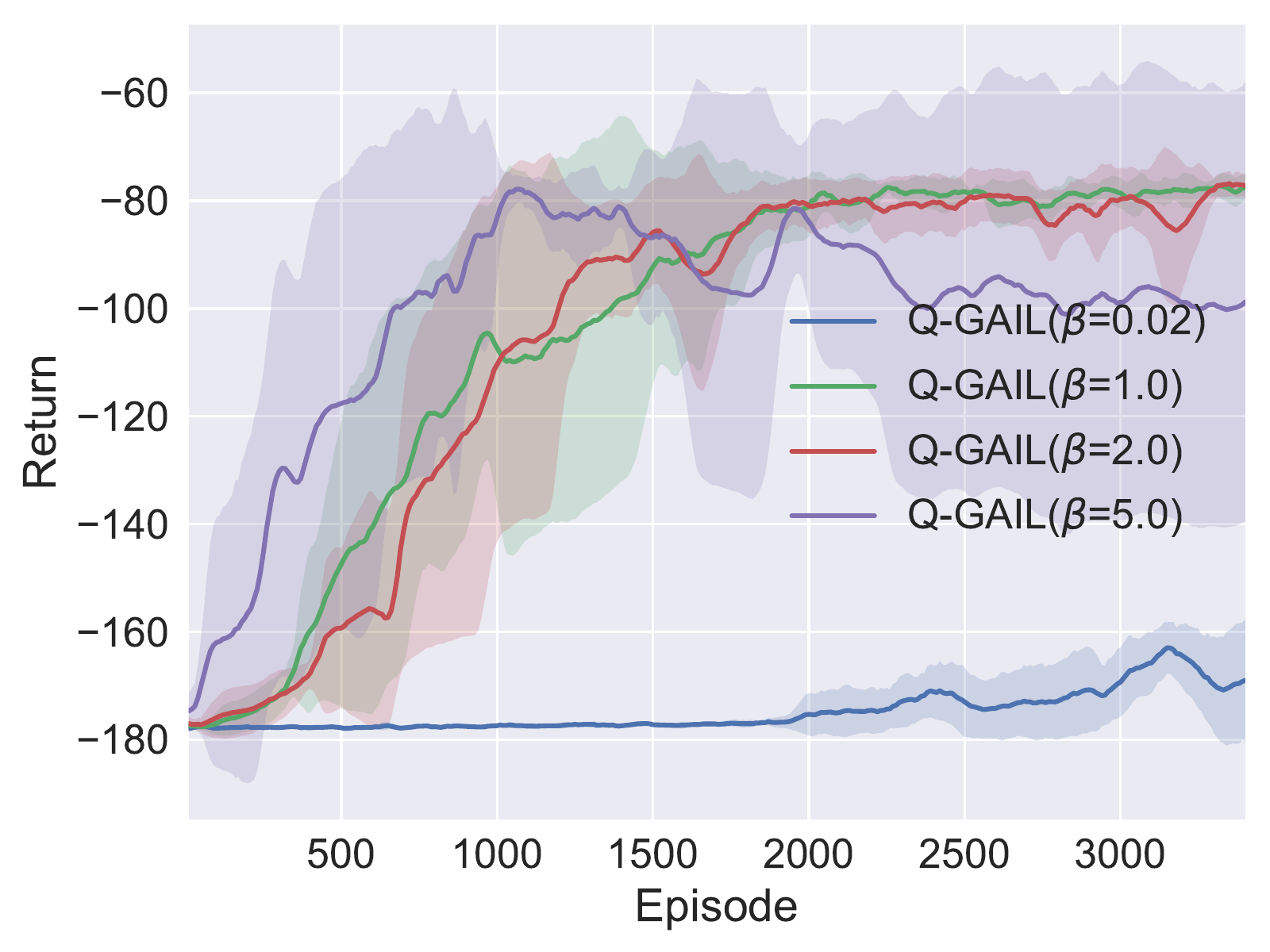}
      }
  \caption{Ablations of the inverse temperature $\beta$.}
  \label{fig:ablationbeta}
\end{figure}
\begin{figure}[!h]
    \centering
    \includegraphics[width=0.4\textwidth]{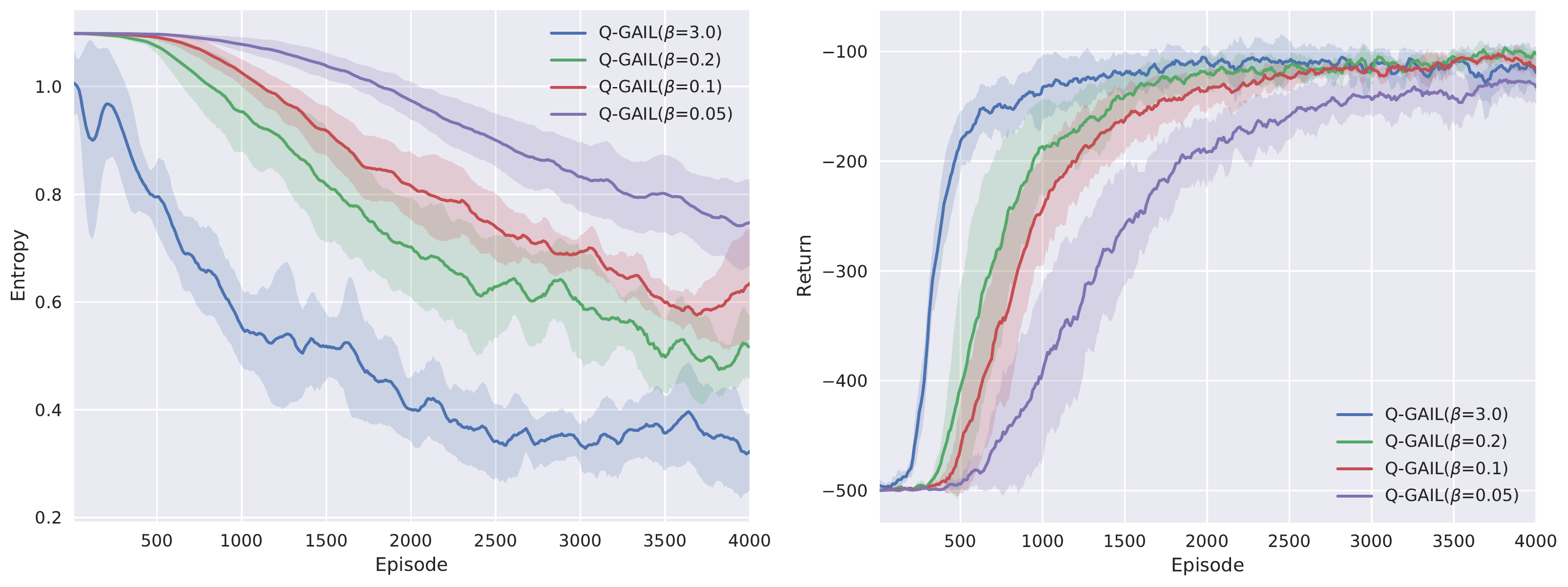}
    \caption{Information entropy of agents with different inverse temperature $\beta$. Left: information entropy, right: return.}
    \label{fig:entropytest}
\end{figure}

\subsubsection{Structures of VQCs}
The softmax-VQC is vital for the success of QIL algorithms. Here, we mainly test the impacts of modifications of softmax-VQC on the IL performance. Concretely, we compare the proposed QIL algorithms with the counterparts whose input and output scaling parameters are disabled, whose policies are constructed with different numbers of layers, and whose $\beta$ is initialized with different values. Besides, we introduce information entropy \cite{shannon2001mathematical} to measure the exploration ability of softmax-VQCs. Experimental results are presented in Figs. \ref{fig:ablationscalingparams}-\ref{fig:entropytest}, respectively. In Figs. \ref{fig:ablationscalingparams}-\ref{fig:ablationbeta}, the first column presents learning curves of Acrobot-v1, and the figures in the second column belong to MountainCar-v0. In Fig. \ref{fig:entropytest}, experiments are conducted in Acrobot-v1 with Q-GAIL. From Fig. \ref{fig:ablationscalingparams}, we can see that both Q-BC and Q-GAIL perform poorly with scaling parameters $\lambda$ and $\nu$ disabled. An interesting phenomenon is that disabling $\lambda$ does not cause damage to the performance of Q-BC in the both two environments. In contrast, Q-GAIL performs poorer with $\lambda$ disabled.

Similarly, in Fig. \ref{fig:ablationlayers}, with fewer layers of softmax-VQC, performance degradations of Q-BC and Q-GAIL are observed. We also notice that the number of layers of softmax-VQC has a greater impact on the performance of Q-GAIL. In MountainCar-v0, Q-BC can successfully reproduce expert behaviors with only one layer. These results mean that enabling scaling parameters and adding more VQC layers help improve the representation ability, leading to a performance gain in IL.

The inverse temperature $\beta$ provides a direct control of exploration of the quantum policy, especially for Q-GAIL. As can be seen from Figs. \ref{fig:ablationbeta} and \ref{fig:entropytest}, a larger $\beta$ generates a more exploitable policy whose entropy is lower, which is prone to finding suboptimal policies, while a smaller $\beta$ could result in a more exploring policy, slowing down the learning process. It is also noticed that the information entropy decreases with training, \emph{i.e.}, the policy learns to perform the task and becomes more deterministic.

\subsubsection{Spectral Normalization}

\begin{figure}[!t]
  \centering
      \subfigure[Acrobot-v1]{
         \includegraphics[width=0.22\textwidth]{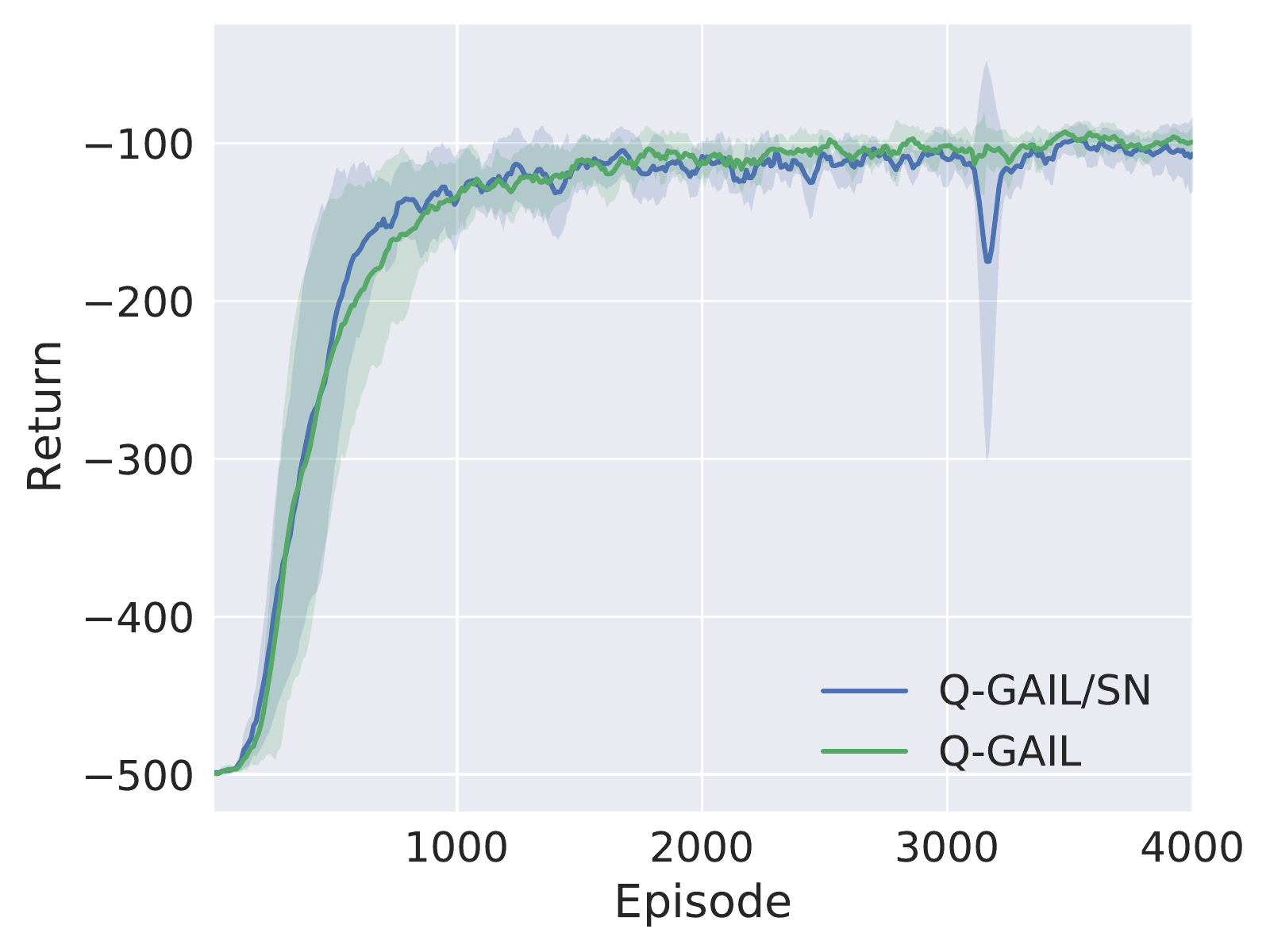}
      }
      \subfigure[MountainCar-v0]{
         \includegraphics[width=0.22\textwidth]{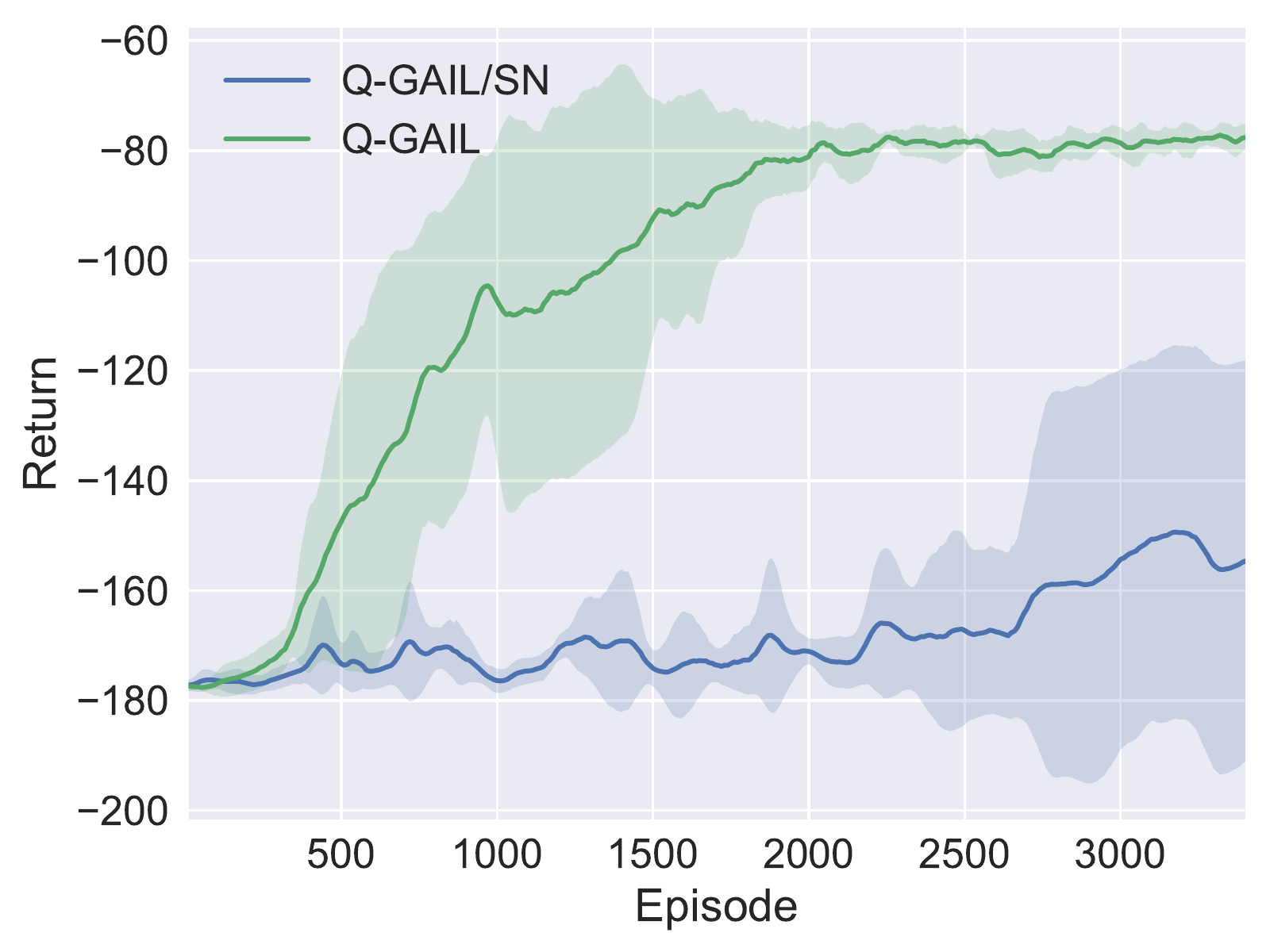}
      }
  \caption{Ablations of spectral normalization. The legend Q-GAIL/SN stands for the Q-GAIL whose spectral normalization is disabled.}
  \label{fig:ablationSN}
\end{figure}

In Q-GAIL, we adopt spectral normalization to further stabilize the adversarial learning scheme. Here, we compare the performance of Q-GAIL with/without spectral normalization in Fig. \ref{fig:ablationSN}. In Acrobot-v1, Q-GAIL performs slightly better than the counterpart Q-GAIL/SN, while Q-GAIL significantly outperforms Q-GAIL/SN in MountainCar-v0. In both environments, Q-GAIL is more stable than the one without spectral normalization. It is clear that spectral normalization plays an important role of improving and stabilizing Q-GAIL. 

\subsubsection{Quantum discriminators}

\begin{figure}[!t]
  \centering
      \subfigure[CartPole-v1]{
         \includegraphics[width=0.2\textwidth]{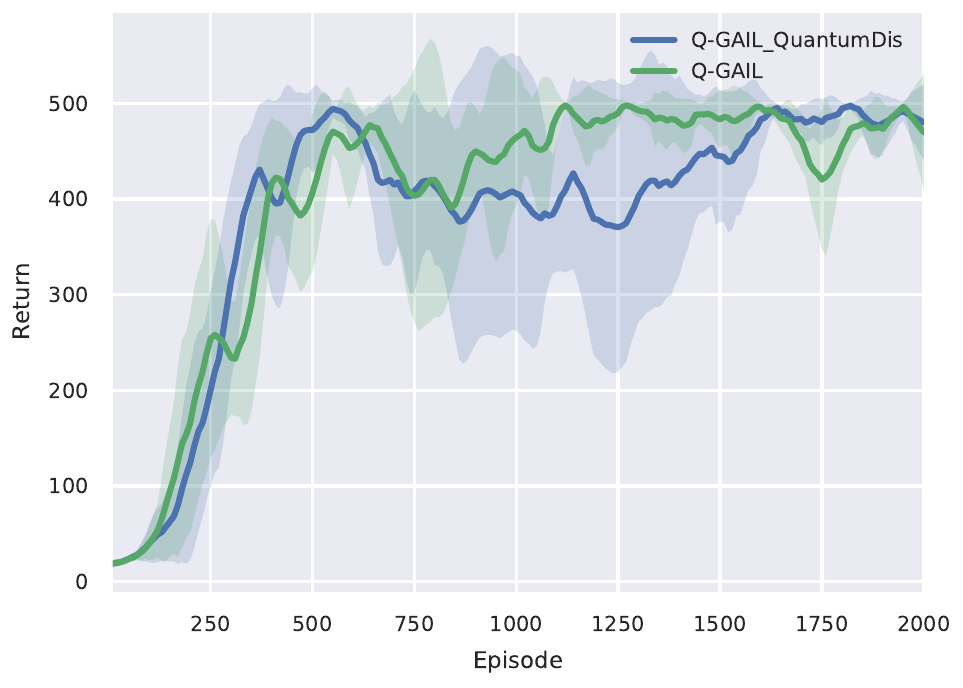}
      }
      \subfigure[Acrobot-v1]{
         \includegraphics[width=0.2\textwidth]{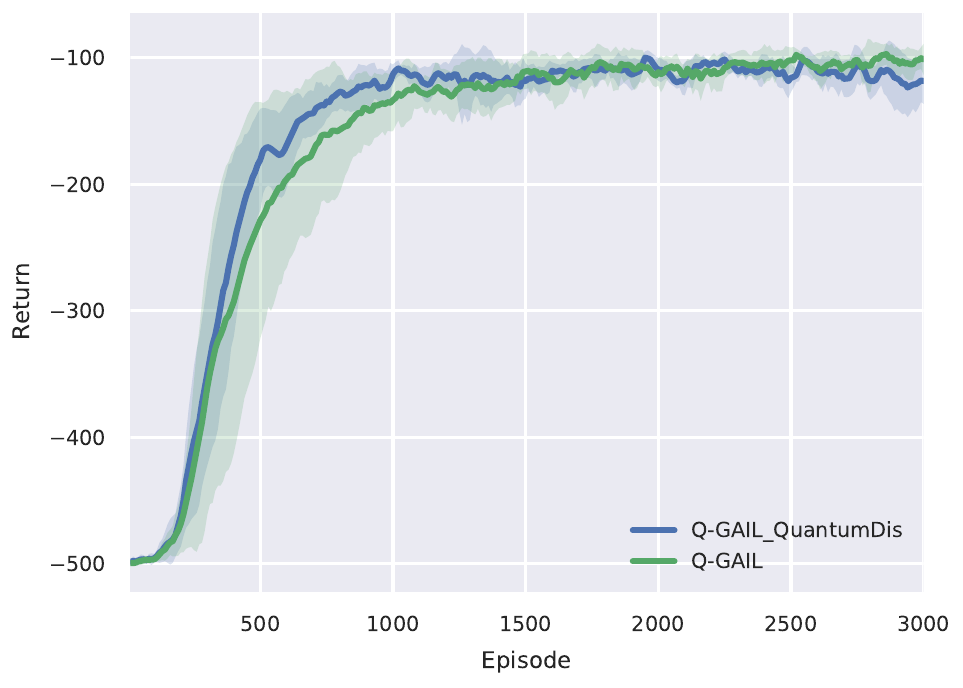}
      }
  \caption{Learning curves of Q-GAIL with a VQC-based discriminator. The legend Q-GAIL\_QuantumDis means that its discriminator is implemented with VQCs.}
  \label{fig:qdis_learning_curves}
\end{figure}

Considering that the discriminator is used to output scalars, we remove the softmax layer at the end of the discriminator. In other words, we directly take the measurement $\bra{0^{\otimes n}} U^{\dagger}(s,\mu) O_a U(s,\mu)\ket{0^{\otimes n}}$ as the discriminator output. In addition, the spectral normalization for DNNs is not adopted for the VQC-based discriminator. We validate the performance of Q-GAIL with the VQC-based discriminator in CartPole-v1 and Acrobot-v1. The learning rate for the parameter $\xi_\theta=[\xi_\lambda, \xi_\phi, \xi_\nu]$ in the discriminator is set to $[0.1, 0.01, 0.1]$. Experimental results are presented in Fig. \ref{fig:qdis_learning_curves}. From Fig. \ref{fig:qdis_learning_curves}, it is clear that Q-GAIL with a VQC-based discriminator performs similarly to the one with a DNN-based discriminator, demonstrating the feasibility to completely abandon DNNs in Q-GAIL.

\subsection{Quantum Advantage}\label{subsec:quantumadvantage}
The aim to conduct QIL is to fuel traditional IL by quantum computing with the potential hope to enjoy the quantum advantage. In the paper, we followed a VQC-based manner to develop QIL algorithms. Although suitable for NISQ devices, VQC-based methods are heuristic and lack theoretical analysis on the quantum speed-up. Our QIL algorithms are possible to run on NISQ computers, being potential to utilize the quantum advantage. One possible reason for speed-up is that the quantum parallelism enables QIL to use quantum policies and actions, which could make the exploration and evaluation of rewards more efficient \cite{agunbiade2022quantum,ren2022nft}. Experiment results show that the proposed QIL approaches achieve better sample efficiency compared to the classical one, especially for Q-GAIL and the continuous task. Besides, fewer trainable parameters are adopted in QIL (\emph{e.g.}, 78 and 4610 trainable parameters are adopted for QIL and IL in CartPole-v1, respectively), indicating the stronger generalization ability~\cite{caro2022generalization}. It should be noted that the improvement in sample efficiency is not satisfactory and far from the potential quadratic or exponential quantum speed-up. It is possible that current problems are small scale such that it is not enough to illustrate the speed-up \cite{buluta2009quantum}. In summary, quantum computing and QIL are still in their infancy and remain an open question that requires further efforts to prove and discover the quantum advantage.

\section{Conclusion}\label{sec:conclusion}
In this paper, we study how to empower IL algorithms with the potential quantum advantage, which paves the way for applications of IL in the coming quantum era. In particular, we employ VQCs with enhanced representation and exploration abilities to replace DNNs. Based on modified VQCs, we propose two QIL algorithms, Q-BC and Q-GAIL. Q-BC mimics experts in an off-line manner, while Q-GAIL needs to interact with environments, which is on-line and on-policy. Both of the two QIL algorithms can mimic experts with expert data and can achieve expert-level performance in discrete-action and continuous-action tasks. 

In the future, we would provide theoretical analyses on the sample complexity of these two QIL algorithms. We would also pay more attention to extending the proposed QIL algorithms to large-scale tasks and diverse fields such as computer vision and text analysis, as well as integrating environments into quantum computers to achieve the potential full speed-up of QIL algorithms.

{\appendix[Proof of Theorem \ref{theorem:bcbound}]
In this section, we provide the detailed proof of Theorem~\ref{theorem:bcbound}. First, we rewrite some necessary lemmas in \cite{xu2021error,jerbi2021parametrized}, which helps simplify the derivation.
\begin{lemma}\label{lemma:TVdistanceoftwooccupancymeasure} (Lemma 4 in \cite{xu2021error})
  Given two policies $\pi$ and $\pi_E$, we obtain that
  \begin{align*}
      D_{TV}(d_\pi, d_{\pi_E})\le\frac{\gamma}{1-\gamma}\mathbb{E}_{s \sim d_{\pi_E}}[D_{TV}(\pi(\cdot |s), \pi_E(\cdot |s))],
  \end{align*}
  where $D_{TV}(d_\pi, d_{\pi_E})$ represents the total variation (TV) distance between $d_\pi$ and $d_{\pi_E}$.
\end{lemma}
\begin{lemma}\label{lemma:stateactiocutostate}(Lemma 5 in \cite{xu2021error})
  Given two policies $\pi$ and $\pi_E$, we get
  \begin{align*}
      D_{TV}(\rho_\pi, \rho_{\pi_E})\le\frac{1}{1-\gamma}\mathbb{E}_{s \sim d_{\pi_E}}[D_{TV}(\pi(\cdot |s), \pi_E(\cdot |s))].
  \end{align*}
\end{lemma}
\begin{lemma}\label{lemma:valuegap}(Lemma 6 in \cite{xu2021error})
  Given two policies $\pi$ and $\pi_E$, we have that
  \begin{align*}
      |J(\pi_E)-J(\pi_\theta)|\le \frac{2R_{MAX}}{1-\gamma}D_{TV}(\rho_{\pi_\theta}, \rho_{\pi_E}).
  \end{align*}
\end{lemma}
\begin{lemma}\label{lemma:TVdisbetweenquantumpolicies}
  Suppose Assumption \ref{assumption:quantummeasureerror} hold, we can obtain the TV distance between the true policy $\pi_\theta$ and the approximated policy $\widetilde \pi_\theta$
  \begin{align*}
      \mathbb{E}_{s \sim d_{\pi_E}}[D_{TV}(\widetilde \pi_\theta(\cdot |s), \pi_\theta(\cdot |s))]\le|\sinh(2\beta \epsilon)|.
  \end{align*}
\end{lemma}
Lemma \ref{lemma:TVdisbetweenquantumpolicies} is used to measure the TV distance between the true policy $\pi_\theta$ and the approximated policy $\widetilde \pi_\theta$.}
\begin{proof}
Recall the definitions of $\pi_\theta$ and $\widetilde \pi_\theta$
 \begin{align*}
     \pi_{\theta}(a|s)&= \frac{e^{\beta \langle O_{a}\rangle_{s,\theta} }}{\sum_{a'}e^{\beta \langle O_{a'}\rangle_{s,\theta}}}\\
     \widetilde \pi_{\theta}(a|s)&= \frac{e^{\beta \langle \widetilde O_{a}\rangle_{s,\theta} }}{\sum_{a'}e^{\beta \langle \widetilde O_{a'}\rangle_{s,\theta}}}.
 \end{align*}
 The quantum measurement error $\epsilon$ in Assumption \ref{assumption:quantummeasureerror} specifies the relationship between $\langle O_a\rangle_{s,\theta}$ and $ \langle \widetilde  O_a\rangle_{s,\theta} $ such that
 \begin{align*}
    \left| \langle \widetilde O_a\rangle_{s,\theta} - \langle O_a\rangle_{s,\theta} \right| \le \epsilon, \forall a \in A.
\end{align*}
According to the monotonicity of the exponential function, we obtain
\begin{align*}
    \frac{e^{\beta \langle \widetilde O_a\rangle_{s,\theta} }}{\sum_{a'}e^{\beta \langle \widetilde O_{a'}\rangle_{s,\theta}}} &\ge \frac{e^{-\beta \epsilon}e^{\beta \langle O_a\rangle_{s,\theta} }}{e^{\beta \epsilon} \sum_{a'}e^{\beta \langle O_{a'}\rangle_{s,\theta}}}= \frac{1}{e^{2\beta \epsilon}}\frac{e^{\beta \langle O_a\rangle_{s,\theta} }}{ \sum_{a'}e^{\beta \langle O_{a'}\rangle_{s,\theta}}} \\
    \frac{e^{\beta \langle \widetilde O_a\rangle_{s,\theta} }}{\sum_{a'}e^{\beta \langle \widetilde O_{a'}\rangle_{s,\theta}}} &\le \frac{e^{\beta \epsilon}e^{\beta \langle O_a\rangle_{s,\theta} }}{e^{-\beta \epsilon} \sum_{a'}e^{\beta \langle O_{a'}\rangle_{s,\theta}}}=e^{2\beta \epsilon}\frac{e^{\beta \langle O_a\rangle_{s,\theta} }}{ \sum_{a'}e^{\beta \langle O_{a'}\rangle_{s,\theta}}}.
\end{align*}
In other words, the approximated policy $\widetilde \pi_{\theta}(a|s)$ is bounded by
\begin{align*}
    e^{-2\beta \epsilon} \pi_{\theta}(a|s) \le \widetilde \pi_{\theta}(a|s) \le  e^{2\beta \epsilon} \pi_{\theta}(a|s).
\end{align*}
Then, the TV distance between $\pi_\theta(\cdot |s)$ and $\widetilde \pi_\theta(\cdot |s)$ with the distribution $s \sim d_{\pi_E}$ can be derived
\begin{align*}
\begin{split}
    \mathbb{E}_{s \sim d_{\pi_E}}&[D_{TV}(\pi_\theta(\cdot |s), \widetilde \pi_\theta(\cdot |s))]\\
    &=\sum_{s}d_{\pi_E}(s)\frac{1}{2}\sum_a|\pi_\theta(a|s) - \widetilde \pi_\theta(a|s)|\\
    &\le \sum_{s}d_{\pi_E}(s)\frac{1}{2}\sum_a|e^{2\beta \epsilon} \pi_\theta(a|s) - e^{-2\beta \epsilon} \pi_\theta(a|s)| \\
    &=\sum_{s}d_{\pi_E}(s)\frac{1}{2}\sum_a|e^{2\beta \epsilon} - e^{-2\beta \epsilon}| \pi_\theta(a|s)\\
    &=\sum_{s}d_{\pi_E}(s)\frac{1}{2}\sum_a 2|\sinh(2\beta \epsilon )|\pi_\theta(a|s)\\
    &=|\sinh(2\beta \epsilon )|.
\end{split}
\end{align*}
\end{proof}

Up to now, we can prove Theorem \ref{theorem:bcbound}.
\begin{proof}
 Based on Lemma \ref{lemma:valuegap}, we have
 \begin{align*}
 \begin{split}
    |J(\pi_E)-J(\pi_\theta)| &\leq \frac{2R_{MAX}}{1-\gamma} D_{TV}(\rho_{\pi_\theta},\rho_{\pi_E})\\
    &\mathop{\leq}^{(i)} \frac{2R_{MAX}}{(1-\gamma)^2}\mathbb{E}_{s \sim d_{\pi_E}}[D_{TV}(\pi_\theta(\cdot |s), \pi_E(\cdot |s))],
 \end{split}
 \end{align*}
 where $(i)$ follows from Lemma \ref{lemma:stateactiocutostate}. And the expectation of the TV distance $\mathbb{E}_{s \sim d_{\pi_E}}[D_{TV}(\pi_\theta(\cdot |s), \pi_E(\cdot |s))]$ can be further simplified
 \begin{align*}
\begin{split}
    \mathbb{E}_{s \sim d_{\pi_E}}&[D_{TV}(\pi_\theta(\cdot |s), \pi_E(\cdot |s))]\\
    =&\frac{1}{2}\sum_{s,a}|\pi_\theta(a|s)-\pi_E(a|s)| \\
    =&\frac{1}{2}\sum_{s,a}|\pi_\theta(a|s)-\widetilde \pi_\theta(a|s) + \widetilde \pi_\theta(a|s)-\pi_E(a|s)|\\
    \le& \frac{1}{2}\sum_{s,a}|\pi_\theta(a|s)-\widetilde \pi_\theta(a|s)| +| \widetilde \pi_\theta(a|s)-\pi_E(a|s)|\\
    =&\mathbb{E}_{s \sim d_{\pi_E}}[D_{TV}(\pi_\theta(\cdot |s), \widetilde \pi_\theta(\cdot |s))]\\
    &+\mathbb{E}_{s \sim d_{\pi_E}}[D_{TV}(\widetilde \pi_\theta(\cdot |s), \pi_E(\cdot |s))].
\end{split}
\end{align*}
In the above equation, the expectation of the first TV distance, $\mathbb{E}_{s \sim d_{\pi_E}}[D_{TV}(\pi_\theta(\cdot |s), \widetilde \pi_\theta(\cdot |s))]$, has been obtained in Lemma \ref{lemma:TVdisbetweenquantumpolicies}. The expectation of the second TV distance, $\mathbb{E}_{s \sim d_{\pi_E}}[D_{TV}(\widetilde \pi_\theta(\cdot |s), \pi_E(\cdot |s))]$, can be proceeded as follows
\begin{align*}
\begin{split}
    \mathbb{E}_{s \sim d_{\pi_E}}&[D_{TV}(\widetilde \pi_\theta(\cdot |s), \pi_E(\cdot |s))]\\
    &\mathop{\le}^{(i)} \mathbb{E}_{s \sim d_{\pi_E}} \sqrt{2[D_{KL}(\widetilde \pi_\theta(\cdot |s), \pi_E(\cdot |s))]}\\
    &\mathop{\le}^{(ii)} \sqrt{2\mathbb{E}_{s \sim d_{\pi_E}}[D_{KL}(\widetilde \pi_\theta(\cdot |s), \pi_E(\cdot |s))]}\\
    &\le \sqrt{2} \sqrt{\delta},
\end{split}
\end{align*}
where $(i)$ follows from the Pinsker’s inequality \cite{csiszar2011information}, and $(ii)$ follows from Jensen’s inequality. Hence, we have that
\begin{align*}
    |J(\pi_E)-J(\pi_\theta)|\leq \frac{2R_{MAX}}{(1-\gamma)^2}(|\sinh(2\beta \epsilon )| + \sqrt{2} \sqrt{\delta}).  
\end{align*}
This completes the proof.
\end{proof}

\bibliographystyle{IEEEtran}
\bibliography{IEEEabrv,mybibfile}

\end{document}